\documentclass[aps,twocolumn,showpacs,superscriptaddress]{revtex4-1}
\usepackage{dsfont,amsthm,amsbsy}
\usepackage{amssymb}
\usepackage{amsmath}
\usepackage{bbm}
\usepackage{graphicx}
\usepackage{epstopdf}
\usepackage{subfigure}
\usepackage{natbib}
\usepackage{epsfig}
\usepackage{amsfonts}
\usepackage{mathrsfs}
\usepackage{sidecap}
\usepackage{lipsum}
\usepackage[toc,page,title,titletoc,header]{appendix}
\usepackage[colorlinks,linkcolor=blue,citecolor=blue,anchorcolor=blue, urlcolor=blue]{hyperref}
\usepackage{hyperref}
\usepackage{resizegather}
\usepackage{tikz}
\usepackage{float}
\usepackage{mathbbol}
\usepackage[normalem]{ulem}
\usepackage{cancel}

\newcommand{\bl}{\begin{aligned}}
\newcommand{\el}{\end{aligned}}
\def\be{\begin{equation}}
\def\ee{\end{equation}}

\def\bi{\begin{itemize}}
\def\ei{\end{itemize}}
\def\bn{\begin{enumerate}}
\def\en{\end{enumerate}}
\def\bea{\begin{eqnarray}}
\def\eea{\end{eqnarray}}

\def\ba{\begin{array}}
\def\ea{\end{array}}
\def\bd{\begin{displaymath}}
\def\ed{\end{displaymath}}

\begin{document}

\title{Quench dynamics and zero-energy modes: the case of the Creutz model}

\author{R. Jafari}
\email[]{jafari@iasbs.ac.ir, rohollah.jafari@gmail.com}
\affiliation{Department of Physics, Institute for Advanced Studies in Basic Sciences (IASBS), Zanjan 45137-66731, Iran}
\affiliation{Beijing Computational Science Research Center, Beijing 100094, China}
\affiliation{Department of Physics, University of Gothenburg, SE 412 96 Gothenburg, Sweden}

\author{Henrik Johannesson}
\email[]{henrik.johannesson@physics.gu.se}
\affiliation{Department of Physics, University of Gothenburg, SE 412 96 Gothenburg, Sweden}
\affiliation{Beijing Computational Science Research Center, Beijing 100094, China}

\author{A. Langari}
\email[]{langari@sharif.edu}
\affiliation{Physics Department, Sharif University of Technology, Tehran 11155-9161, Iran}

\author{M. A. Martin-Delgado}
\email[]{mardel@miranda.fis.ucm.es}
\affiliation{Departamento de Fi­sica Te\'orica I, Universidad Complutense, 28040 Madrid, Spain}
\affiliation{CCS -Center for Computational Simulation, Campus de Montegancedo UPM, 28660 Boadilla del Monte, Madrid, Spain}

\date{\today}

\begin{abstract}
In most lattice models, the closing of a band gap typically occurs
at high-symmetry points in the Brillouin zone. Differently, in the Creutz model $-$ describing a system of spinless fermions hopping on
a two-leg ladder pierced by a magnetic field $-$
the gap closing at the quantum phase transition between the two topologically nontrivial phases of the model can be
moved by tuning the hopping amplitudes.
We take advantage of this property
to examine the nonequilibrium dynamics of the model
after a sudden quench of the magnetic flux through the plaquettes of the ladder. 
For a quench to one of the equilibrium quantum critical points we find that the revival period of the Loschmidt
echo
$-$ measuring the overlap between initial and time-evolved states $-$
is controlled by the gap closing zero-energy modes.
In particular, and contrary to expectations, the revival period of the Loschmidt echo for a finite ladder does not scale linearly with size but exhibits jumps determined
by the presence or absence of zero-energy modes.
We further investigate the conditions for the appearance of dynamical quantum phase
transitions in the model and find that, for a quench {\em to} an equilibrium critical point, such transitions occur only for ladders of sizes which host zero-energy modes.
Exploiting concepts from quantum thermodynamics, we show that the average work and the irreversible work per lattice site exhibit a weak dependence on the size of
the system after a quench {\em across} an equilibrium critical point, suggesting that quenching into a different phase induces effective correlations among the particles.

\end{abstract}

\maketitle

\section{Introduction \label{introduction}}
Recent progress in the studies of ultra-cold atoms trapped in optical lattices provide a new
framework for investigating the nonequilibrium dynamics of quantum critical phenomena \cite{Bloch,Polkovnikov2011,Dziarmaga}.
While there are many ways to drive a physical system out of equilibrium,
the simplest controllable scheme is arguably that of a quantum quench.
Here a system is prepared in a well-defined initial state and then taken out of equilibrium by a
change of a Hamiltonian parameter  \cite{Greiner2002,Calabrese2006} or by a projective measurement \cite{Bayat2018}.
The nonequilibrium dynamics of a quenched quantum system can be described and characterized in many different ways.
In the case of a sudden quench, a very efficient approach is to employ the notion of
the Loschmidt echo (LE) \cite{Gorin} $-$ the modulus of the Loschmidt amplitude (LA) $-$ which measures the overlap of
the initial quantum state with its time-evolved state controlled by the post-quench Hamiltonian.
In fact, the LE has been explored for a variety of problems connected directly or indirectly
to quench dynamics, including
quantum chaos \cite{Peres,Jalabert,Huang},
quantum speed limit time \cite{Zou}, quantum decoherence \cite{Yuan,Quan,Cucchietti2007,Mostame,Sun2007,Jafari2015,Pazbook}, equilibrium quantum phase transitions \cite{Quan,Liu,Montes,Happola,Heyl2013,Dorner,Haikka,Jafari2016,Kolodrubetz}, dynamical quantum phase transitions \cite{Heyl2013,Heyl2015,Jurcevic,Flaschner,Karraschb,Sedlmayr,Ning,Zache,Jada,Jadc,Jadd,Jade,Jadb},
work statistics \cite{Silva,Kolodrubetz,Campbell} and entropy production \cite{Dorner}.

Concentrating on quantum criticality, a central problem has been to link the salient features of quench dynamics to equilibrium quantum
phase transitions (QPTs) \cite{Quan,Montes,Happola,Heyl2013,Campbell,Dorner,Karrasch}.
The LE has here been used to pinpoint how distinct signatures of an equilibrium
QPT are manifested in the dynamics when a system is quenched {\em to} a quantum critical point \cite{Quan,Montes,Happola,Najafi} as compared
to a quench {\em across} a quantum critical point \cite{Heyl2013,Dorner,Campbell,Kolodrubetz}. An early analytical result for the dynamics
of the one-dimensional transverse field Ising model \cite{Quan}
suggested that the LE characteristically exhibits an accelerated decay followed by periodic revivals when quenched to a critical point \cite{Montes,Happola} $-$
a finding later noted also for other models \cite{Montes,Happola}.
However, more recent studies show that a periodic Loschmidt revival may or may not appear for this case. What matters is
that the specific modes which contribute to the LE are massless, a property, which is not guaranteed to materialize at a QPT \cite{RJHJ2017a,RJHJ2017b}.

The LE has also turned out to be useful for identifying nonanalyticities in the time evolution of a system out of equilibrium
$-$ a.k.a. a dynamical quantum phase transition (DQPT) \cite{Zvyagin2016,HeylReview,Hickey}. Important recent results \cite{Heyl2013} suggested that such nonanalyticities,
calculated from the LE, are generically linked to a quench across an equilibrium quantum critical point.  Subsequent studies, also employing the concept of an LE,
have revealed that a quench across a critical point does not necessarily imply
a DQPT, and that such a transition may instead occur when quenching {\em to} the critical point {\em within} the same phase of the system \cite{Sharma2015,Vajna,Andraschko}.

Yet another application of the LE to the problem of quantum critical dynamics has been to extract the work distribution function of a system \cite{Silva}.
Notably, it has been shown that the irreversible work and irreversible entropy production signal the presence of a QPT \cite{Silva,Kolodrubetz,Campbell}.
Recently the irreversible work was found to lay bare also the critical properties of
quantum impurity QPTs \cite{Bayat}.

Despite numerous attempts to provide a bridge between QPTs and the quench dynamics encoded in the LE,
a general principle joining the two notions is still missing. To make progress, more studies are needed
so as to obtain a ``critical mass" of results from which a theory can be built. Exactly solvable models here play a
particularly important role.

In this article we try to contribute to this program by studying the quench dynamics of the exactly solvable Creutz model \cite{Creutz} $-$
describing a system of spinless fermions hopping on a two-leg ladder pierced by a magnetic field $-$ using the concept of the LE as a main tool.
As the magnetic field is varied, the model exhibits a quantum phase transition between two insulating phases
with topologically distinct configurations of the induced local charge current \cite{Bermudez}. Depending on the choice of hopping amplitudes for the fermions, the insulating gap
may close at the quantum critical point already for a finite ladder provided that its number of sites is commensurate with the wave number defining the gap closing point, as determined by
the finite-size quantization condition.
Moreover, the location in the Brillouin zone of the associated zero-energy quasiparticle excitations can be moved by tuning the hopping amplitudes \cite{Bermudez}. This is reminiscent of systems with movable accidental or symmetry-enforced spectral degeneracies protected by a local \cite{Murakami} or global \cite{Schnyder} topological charge, respectively.
We take advantage of this property to explore how the gap-closing zero-energy modes govern the quench dynamics of a finite-size Creutz ladder
when quenched to a critical point. Specifically, by changing the hopping amplitudes, we can pinpoint how the zero-energy gap-closing modes control the revival period of
the Loschmidt echo. If these modes are not present in a given finite-size ladder, the revival period is instead determined by that of the nearest commensurate ladder, which contains these modes
(with the precise notion of ``commensurate" to be detailed below). Different from results obtained for other models \cite{Happola,Montes,Cardy,Michal}, this implies that the revival period for a finite Creutz ladder does not scale linearly with size but exhibits jumps determined by the presence or absence of zero-energy modes.
Carrying over our results to the quench dynamics of the transverse field Ising chain explains the intriguing period doubling of the Loschmidt echo revivals reported by H\"app\"ol\"a {\em et al.} \cite{Happola} when the model is subject to periodic boundary conditions.
To emphasize the important role of zero-energy modes also in DQPTs, we analyze a quench to an equilibrium quantum critical point of the Creutz ladder and find that a DQPT in this case happens only if there are zero-energy modes present. We expect this result to hold quite generally.
Concentrating on the case of a sudden magnetic flux quench, we also use concepts from
quantum thermodynamics \cite{Campisi2011} to investigate how the work statistics play out when quenching {\em to} a critical point as compared to quenching {\em across} the same point.
The fact that the quantum critical points that we probe define topological phase transitions adds to the interest of our analysis.

The article is organized as follows. In Sec. \ref{Creutz} we present the model and review its exact solution.
Sec. \ref{LE} is dedicated to an analysis of the LE of the model and the periodic revival structure for a quench to the critical point.
In Sec. \ref{DQPT} the appearance of a dynamical quantum phase transition in the model is explored for both a quench to one of the equilibrium
quantum critical points and a quench across the same point. In Sec. \ref{Work} we examine the average work and the
irreversible work performed on the system by a quench. Sec. \ref{Conclusions} contains some concluding remarks.
%
%
\section{Creutz model \label{Creutz}}
%
\begin{figure}
\includegraphics[width=\columnwidth]{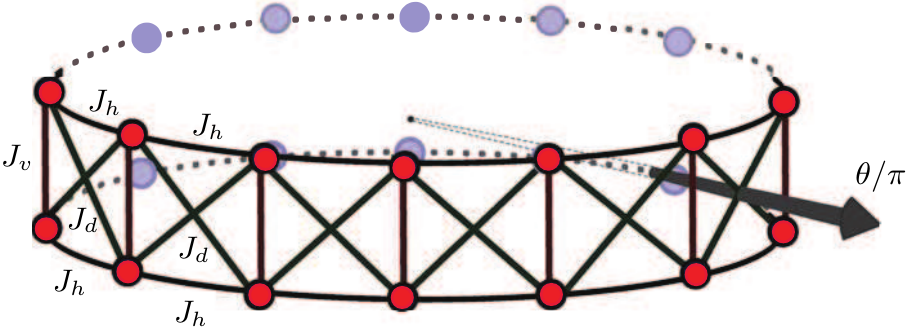}
\caption{ (Color online) The Creutz ladder with
periodic boundary conditions and magnetic flux $\theta/\pi$ per plaquette. Horisontal,
vertical, and diagonal bonds are marked with the corresponding hopping amplitudes
$J_h, J_v$, and $J_d$ respectively.}
\label{fig1}
\end{figure}
%
The Creutz model describes the dynamics of a system of spinless fermions on a two-leg ladder,
depicted in Fig.~\ref{fig1}, and governed by the Hamiltonian \cite{Creutz}
%
\begin{align}
\begin{split}
H =-\sum_{n=1}^{N} & \Big(J_{h}(e^{i\theta}c^{p\dagger}_{n+1}c^{p}_n+e^{-i\theta}c^{q\dagger}_{n+1}c^{q}_n) \\
& + J_{d}\,(c^{p\dagger}_{n+1}c^{q}_n+c^{q\dagger}_{n+1}c^{p}_n)+J_{v} c^{q\dagger}_{n}c^{p}_n \Big) + \mbox{H.c}.
\label{eq1}
\end{split}
\end{align}
%
Here $p$ ($q$) labels the lower (upper) leg of the ladder, with $c^{p/q \dagger}_{m}$ and $c^{p/q}_m$ the corresponding fermion creation-
and annihilation operators respectively, and with $c_{N+1}^{p/q} = c_1^{p/q}$ given periodic boundary conditions.
The magnitude of the hopping amplitudes for horizontal bonds are assumed to be the same for the lower and upper legs and denoted by $J_{h}$.
Similarly, for vertical (diagonal) bonds (cf. Fig. 1), the uniform hopping amplitude is $J_{v} \, (J_{d})$, both taken to be positive and real.
The presence of the gauge-dependent Peierls-type complex phases in (\ref{eq1}), here attached to hopping along the legs of the ladder, mimics the presence
of a magnetic field which pierces the ladder and supplies a magnetic flux $\theta/\pi$ per plaquette (in natural units, cf. Fig. 1).

Introducing the spinor $\Gamma^{\dagger}\!=\!(c^{q\dagger}_{k} c^{p\dagger}_{k})$,
the Fourier transformed Hamiltonian can be expressed 
as $H= \sum_{k\ge0}\Gamma^{\dagger}H(k)\Gamma$, with
%
\begin{figure}
\includegraphics[width=\columnwidth]{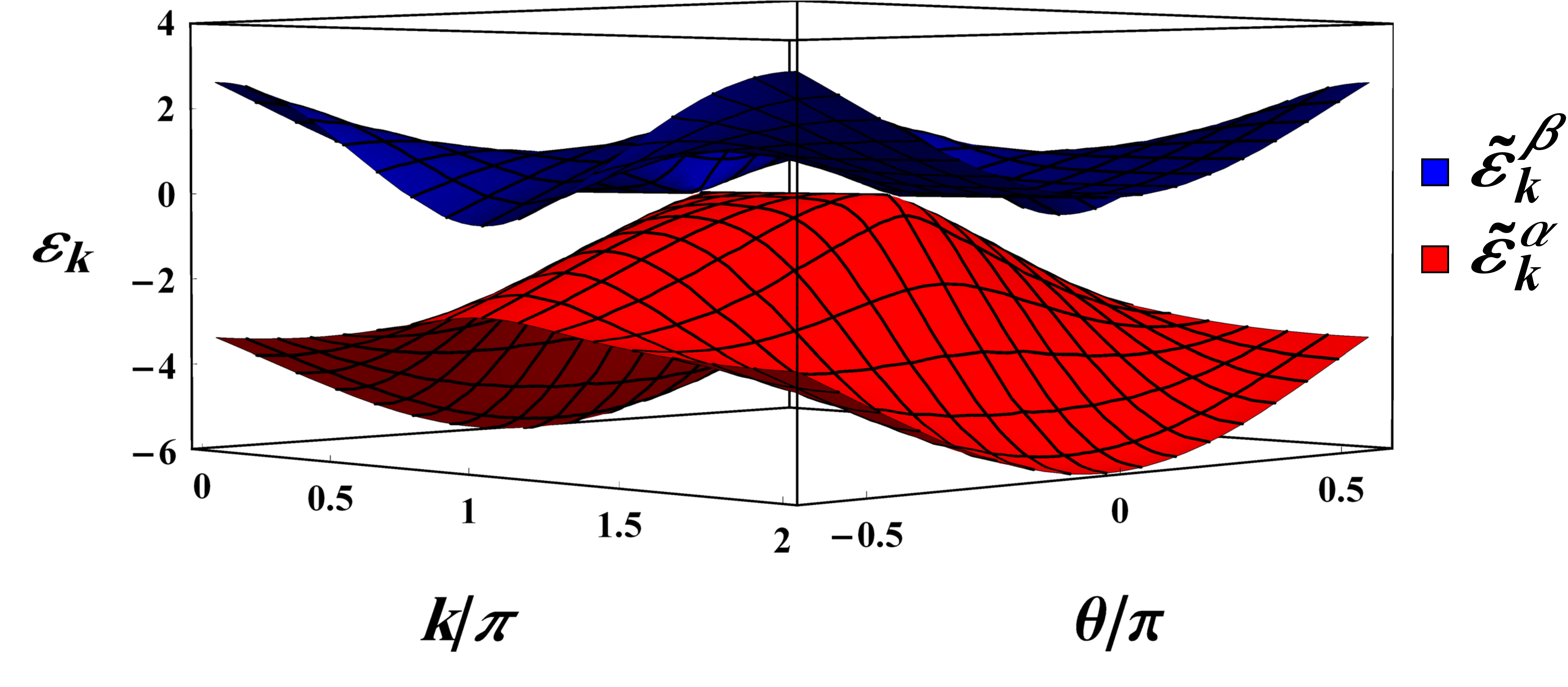}
\caption{ (Color online) Quasiparticle spectra
$\tilde{\varepsilon}^{\alpha}_{k}$ and $\tilde{\varepsilon}^{\beta}_{k}$ versus wave number $k$ and flux $\theta/\pi$ for $J=J_v=1$.}
\label{fig2}
\end{figure}
%
%
\bea
\label{eq2}
H(k)=-
\left(
  \begin{array}{cc}
    \varepsilon_{k}^{q} & \varepsilon_{k}^{qp}  \\
    \\
    \varepsilon_{k}^{qp} &  \varepsilon_{k}^{p} \\
  \end{array}
\right),
\eea
%
where $\varepsilon_{k}^{q}=2J_{h}\cos(k-\theta)$, $\varepsilon_{k}^{p}=2J_{h}\cos(k+\theta)$, and $\varepsilon_{k}^{qp}=2J_{d}\cos(k)+J_{v}$.
Here $k$ is quantized, taking values $k\!=\!k_{j}\!=\!2\pi j/N \ \mbox{with} \ j\!=\!0, ... ,N-1$.
Using a Bogoliubov transformation \cite{Zhu2016},
%
\begin{align}
\begin{split}
\label{eq3}
c_{k}^{q}&=\cos(\gamma_{k}/2) \alpha_{k}+\sin(\gamma_{k}/2) \beta_{k},\\
c_{k}^{p}&=-\sin(\gamma_{k}/2) \alpha_{k}+ \cos(\gamma_{k}/2) \beta_{k},
\end{split}
\end{align}
%
with
%
\bea
\label{eq4}
\tan(\gamma_{k})\!=\!2\varepsilon_k^{qp}/(\varepsilon_k^{q}-\varepsilon_k^{p}),
\eea
%
and with $\alpha_{k}$ and $\beta_{k}$ quasiparticle operators, we can then write the Hamiltonian on diagonalized form,
$H\!=\!\sum_{k}(\varepsilon^{\alpha}_{k}\alpha^{\dagger}_{k} \alpha_{k}+\varepsilon^{\beta}_{k}\beta^{\dagger}_{k}\beta_{k})$,
where
%
\begin{align}
\begin{split}
\label{eq5}
\varepsilon^{\alpha}_{k}(\theta)\!=\!-2J_{h}\!\cos(k)\!\cos(\theta)\!-\!\sqrt{\!(\varepsilon_{k}^{qp})^{2}\!+\!(2J_{h}\!\sin(k)\sin(\theta))^{2}}\\
\varepsilon^{\beta}_{k}(\theta)\!=\!-2J_{h}\!\cos(k)\!\cos(\theta)\!+\!\sqrt{\!(\varepsilon_{k}^{qp})^{2}\!+\!(2J_{h}\!\sin(k)\sin(\theta))^{2}}
\end{split}
\end{align}
%
with corresponding quasiparticle eigenstates
%
\begin{align}
\begin{split}
\label{eq6}
\alpha_{k}^{\dagger}|V\rangle&=\cos(\frac{\gamma_{k}}{2})c_{k}^{q\dagger}|0\rangle
-\sin(\frac{\gamma_{k}}{2})c_{k}^{p\dagger}|0\rangle,\\
\beta_{k}^{\dagger}|V\rangle&=\sin(\frac{\gamma_{k}}{2})c_{k}^{q\dagger}|0\rangle
+\cos(\frac{\gamma_{k}}{2})c_{k}^{p\dagger}|0\rangle,
\end{split}
\end{align}
where $|V\rangle$ and $|0\rangle$ are vacuum states of the quasiparticle and fermion respectively.
In what follows we restrict our analysis to the case of $J_{h}\!=\!J_{d}\!=\!J$,
with redefined quasiparticle energies
$\tilde{\varepsilon}^{\alpha/\beta}_{k}\!\equiv \!\varepsilon^{\alpha/\beta}_{k}\!-\!J_{v}$ (see Fig. \ref{fig2}).

For vertical hopping $J_{v}<2J$, the model is known to have second order quantum phase
transitions at $\theta=\theta_c=0$, $\pi$ \cite{Creutz,Bermudez}, with the band gap $\Delta \tilde{\varepsilon}_{k}(\theta_c)=
\tilde{\varepsilon}^{\beta}_{k}(\theta_c)-\tilde{\varepsilon}^{\alpha}_{k}(\theta_c)$
closing at wave numbers $k^{\pm}_{c}=\pi \pm \arccos(J_{v}/2J)$. Considering the
quantization condition on $k$, and choosing values of $J_v$ and $J$ such that $\arccos(J_v/2J) = (p/q)\pi$ with $p/q \in Q^+$, the vanishing of the gap
for a finite system is seen to require that the number of sites $N$ on each leg of the ladder is a multiple of $2q/(q-p)$ and $2q/(q+p)$, i.e.
\begin{equation} \label{Ncondition}
N = \frac{2q}{q\!\pm\!p}\, m^{\pm}, \ \ \ m^{\pm} \in \mathbb{N}.
\end{equation}
If these conditions are not satisfied when $\arccos(J_v/2J) = (p/q)\pi$, the gap closes only asymptotically in the thermodynamic limit $N \rightarrow \infty$.
Obviously, the distinction between the two cases is immaterial in an experimental realization of the Creutz model with large $N$. However,
as we shall see, it holds the key to understanding the general LE revival structure $-$ and with that, the quantum dynamics $-$ after a
sudden quench to one of the quantum critical points $\theta_c\!=\!0,\pi$. To uncover the full revival structure we shall exploit an expedient
feature of the Creutz model: the movability of the gap-closing modes in the Brillouin zone, controllable by tuning the hopping amplitudes $J_v$ or $J.$

Before concluding this section, let us briefly recall some basic facts about the ground state phase diagram of the Creutz model. The critical
points $\theta_c = 0, \pi$ for $J_v<2J$ separates two topologically nontrivial insulating phases characterized by a Zak phase $\gamma \!=\! \pi\, \mbox{mod} \,2\pi$,
and with opposite circulations of the local charge current around a lattice plaquette \cite{Li2015}.
When $\theta = \pm \pi/2$, there is an emergent chiral symmetry of the model, which,
considering the broken time-reversal symmetry from the magnetic flux, puts the system in the AIII Altland-Zirnbauer symmetry class \cite{Bermudez,Delgado2014,Chiu2016}.
Cutting open the ladder, the topologically nontrivial phases support zero-energy {\em boundary} states at its edges (``zero modes", not to be mixed up with the
gap-closing zero-energy {\em bulk} modes discussed in this article). Importantly, the inversion
symmetry present for any value of $\theta$ protects the topological phases also when chiral symmetry is absent \cite{Chiu2013}.
A topological phase transition occurs at $J_v=2J$ for any value of the flux $\theta/\pi$, with the insulating phase for $J_v>2J$ being
topologically trivial, with $\gamma=0$ \cite{Bermudez}.

%
%
\section{Loschmidt Echo Revivals\label{LE}}
A sudden quench of a quantum system is conventionally carried out by instantaneously changing a parameter in its Hamiltonian $H(\theta_1)$,
with $\theta_1$ denoting the value(s) of the parameter(s) to be changed. (For an alternative protocol, a {\em measurement quench}, see Ref. \onlinecite{Bayat2018}.)
In the case of the Creutz model, $\theta_1$ can be taken as the
Peierls phase appearing in the horizontal hopping amplitudes in Eq. (\ref{eq1}), representing the $\theta_1/\pi$ magnetic flux through a square plaquette of the Creutz ladder (Fig. 1).
If the system is initially prepared in an eigenstate $|\Psi_{m}(\theta_1)\rangle$ of $H(\theta_1)$ and $\theta_1$ is suddenly changed to $\theta_2$ at time $t=0$,
the time evolution of the system becomes governed by the post-quench Hamiltonian $H(\theta_2)$ according to
$|\Psi_{m}(\theta_1, \theta_2,t)\rangle=\exp(-iH(\theta_2)t) |\Psi_{m}(\theta_1)\rangle$.
%
\begin{figure}
\includegraphics[width=\columnwidth]{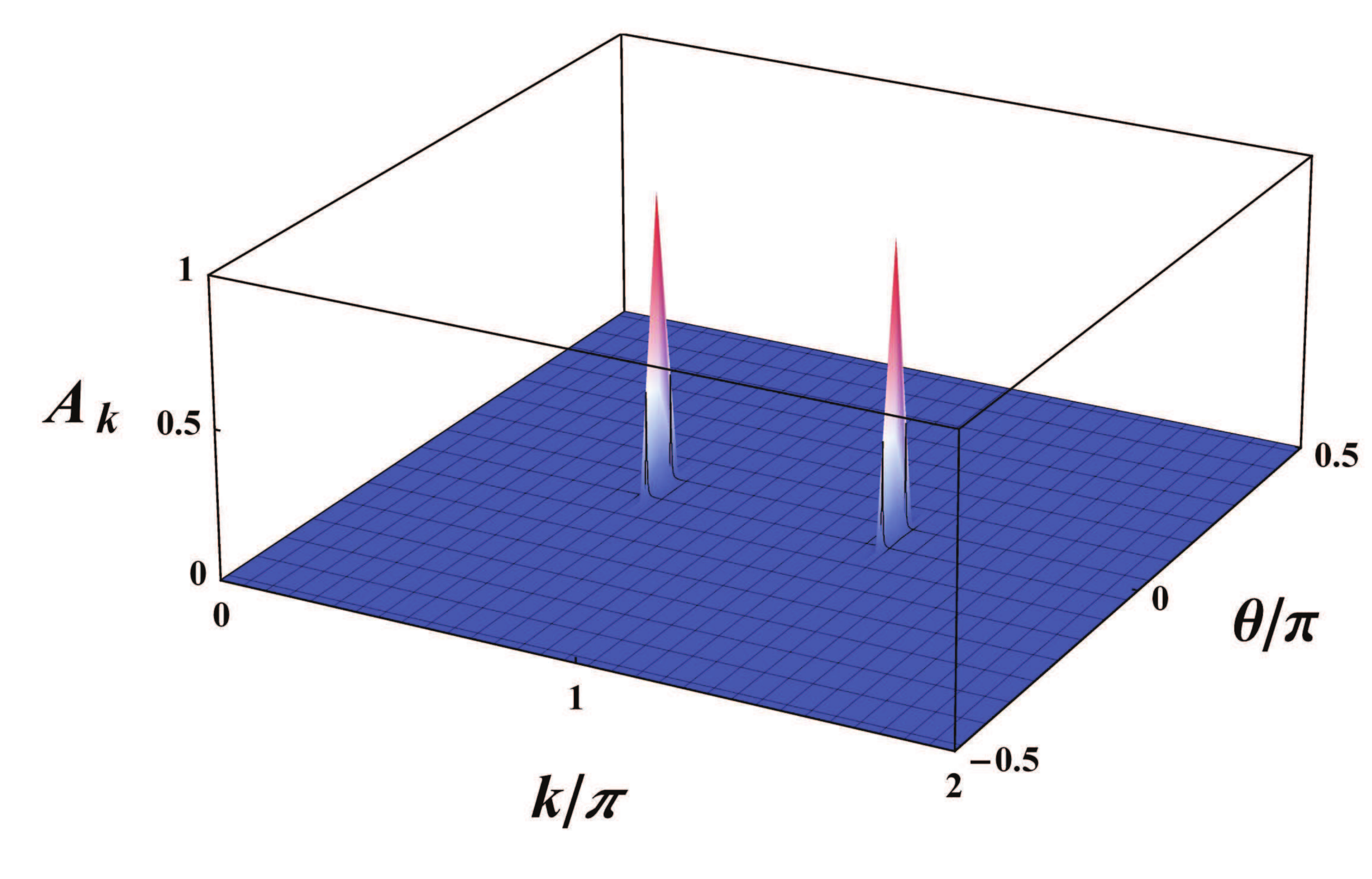}
\caption{ (Color online) Oscillation amplitude
$A_{k}$ in the mode decomposition of the LE, Eq. (\ref{eq8}), versus $k/\pi$ and $\theta/\pi$ when
$J=J_v=1$.}
\label{fig3}
\end{figure}
%
Choosing the initial state as the ground state of the system, call it $|\Psi_{G}(\theta_1)\rangle$, the LE \cite{Gorin}
takes the form of a return probability,
%
\bea
\label{eq7}
{\cal L} (\theta_1,\theta_2,t)= |\langle\Psi_{G}(\theta_1)|\exp(-iH(\theta_2)t) |\Psi_{G}(\theta_1)\rangle|^2,
\eea
%
and can be interpreted as a dynamical version of the ground-state fidelity, providing a measure of the distance between the time-evolved state
$|\Psi_{G}(\theta_1,\theta_2,t)\rangle$ and the initial state $|\Psi_{G}(\theta_1)\rangle$.

To calculate the LE for the Creutz model we imagine that the system is initially prepared in the ground state $|\Psi_{G}(\theta_1)\rangle$,
obtained by filling up the negative-energy quasiparticle states (cf. Eqs. (\ref{eq5}) and (\ref{eq6})), $|\Psi_{G}(\theta_1)\rangle=\prod_{k}\alpha^{\dagger}_{k}|V\rangle$,
assuming that the Fermi level is chosen at zero energy.
A straightforward but lengthy calculation yields the complete set of eigenstates of the model from which an expression for the LE can
be extracted.  Quenching the Peierls phase from $\theta_1$ to $\theta_2$ at $t\!=\!0$ one obtains
\bea
\label{eq8}
{\cal L} (\theta_1,\theta_2,t)\!=\!\prod_{k}\!{\cal L}_k(\theta_1,\theta_2,t)\!=\!\prod_{k}[1\!-\!A_{k}\!\sin^{2}(\frac{\Delta\tilde{\varepsilon}_{k}t}{2})]
\eea
%
where
%
\begin{align}
\begin{split}
\label{eq9}
&\Delta \tilde{\varepsilon}_{k}=\tilde{\varepsilon}^{\beta}_{k}(\theta_2)-\tilde{\varepsilon}^{\alpha}_{k}(\theta_2)
\!=\!2\sqrt{(\varepsilon_{k}^{qp})^{2}+(2J_{h}\sin(k)\sin(\theta_2))^{2}},\\
&A_{k}=\sin^{2}(2\eta_{k}),~2\eta_{k}=\gamma_{k}(\theta_{1})-\gamma_{k}(\theta_{2}).
\end{split}
\end{align}
%
It is worth mentioning that if we had instead considered $\prod_{k}\beta^{(0)}_{k}|V\rangle$ as the initial state of the system, the LE would still have
been governed by Eq. (\ref{eq8}).

The LE decays in a time $T_{\text{rel}}$ {\em (relaxation time)} from unity to an average value around which it then
fluctuates \cite{Venuti2011}. Revivals show up in the LE as distinct deviations
from the average value, forming local maxima \cite{Happola}.
When quenching to a quantum critical point in a finite system there is an expectation
that the LE relaxation rate becomes faster compared with a noncritical quench \cite{Quan,Yuan,Zhang,Rossini2007a,Rossini2007b,Sharma2012,Sacramento}
and that the revivals show a periodicity \cite{Quan,Yuan,Happola,Montes}. Studies have also found a linear scaling of the periodicity of revivals with system size for both sudden \cite{Happola,Montes,Cardy} and slow \cite{Michal} quenches. In fact, a periodic revival structure has frequently been used as a diagnostic of criticality, following
results in Refs.  [\onlinecite{Quan}] and [\onlinecite{Yuan}].
However, recent work has revealed that a quench to a quantum critical point is neither a necessary nor a sufficient condition for periodic revivals \cite{RJHJ2017a,RJHJ2017b}.
%
\begin{figure*}
\centerline{\includegraphics[width=0.33\linewidth]{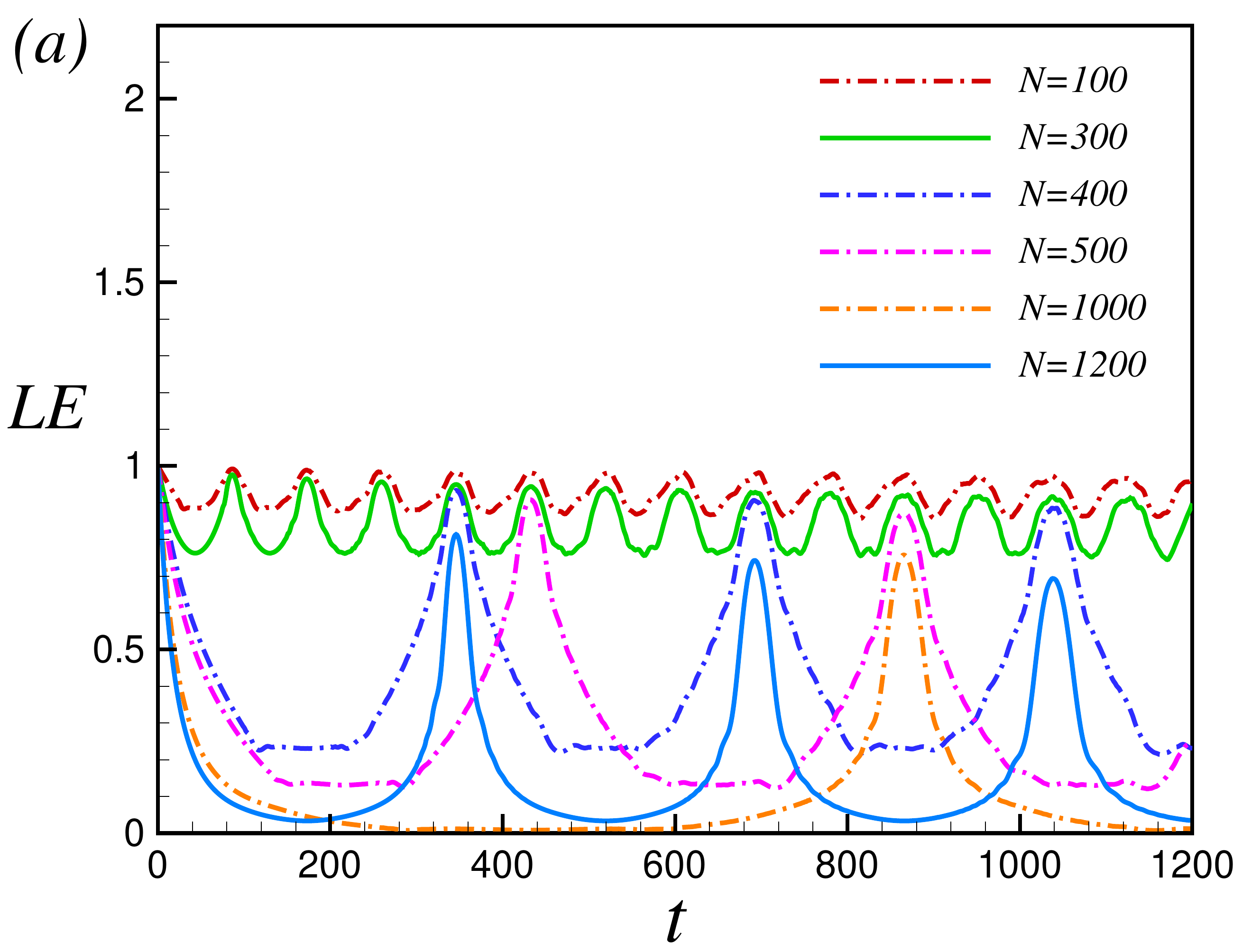}
\includegraphics[width=0.33\linewidth]{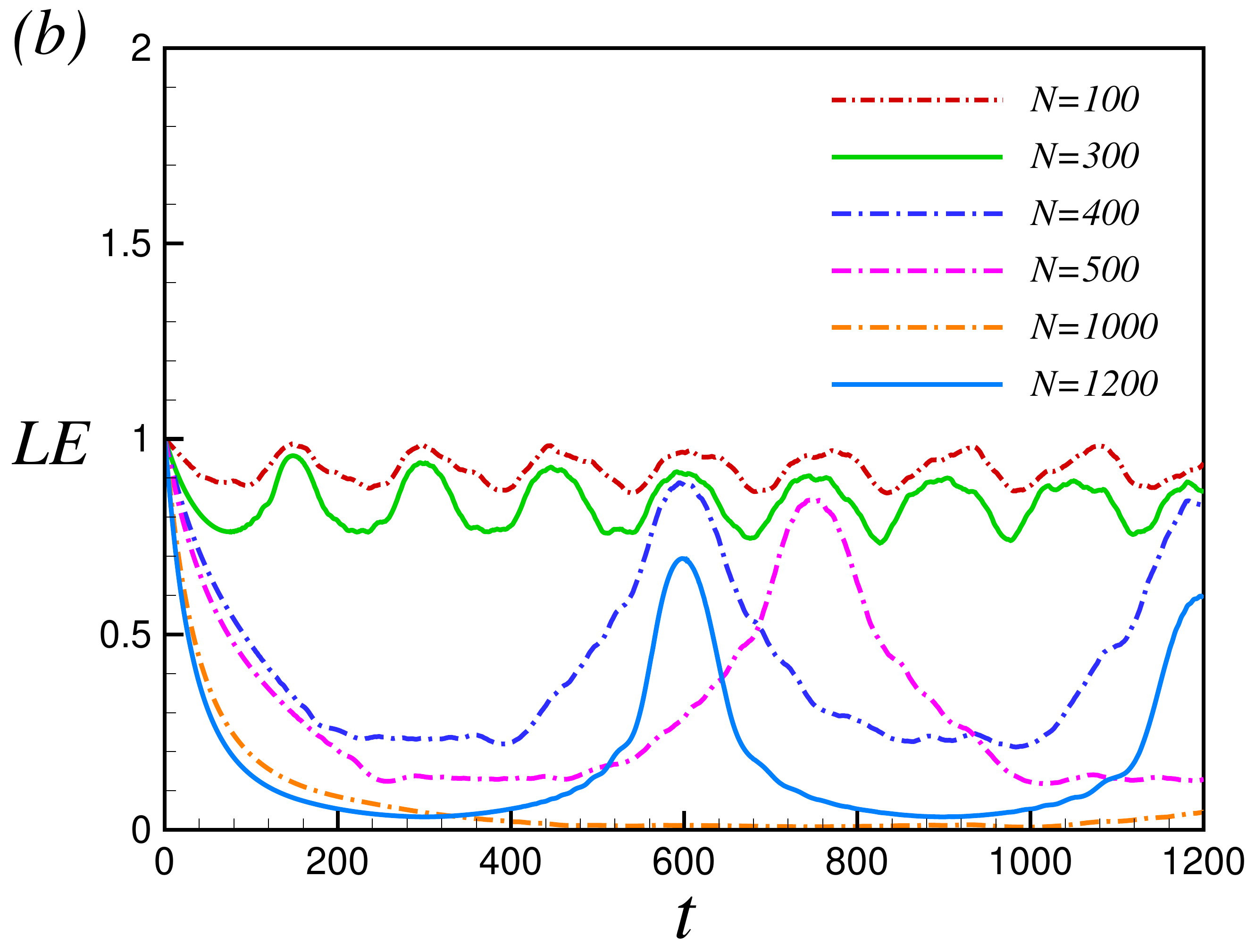}
\includegraphics[width=0.33\linewidth]{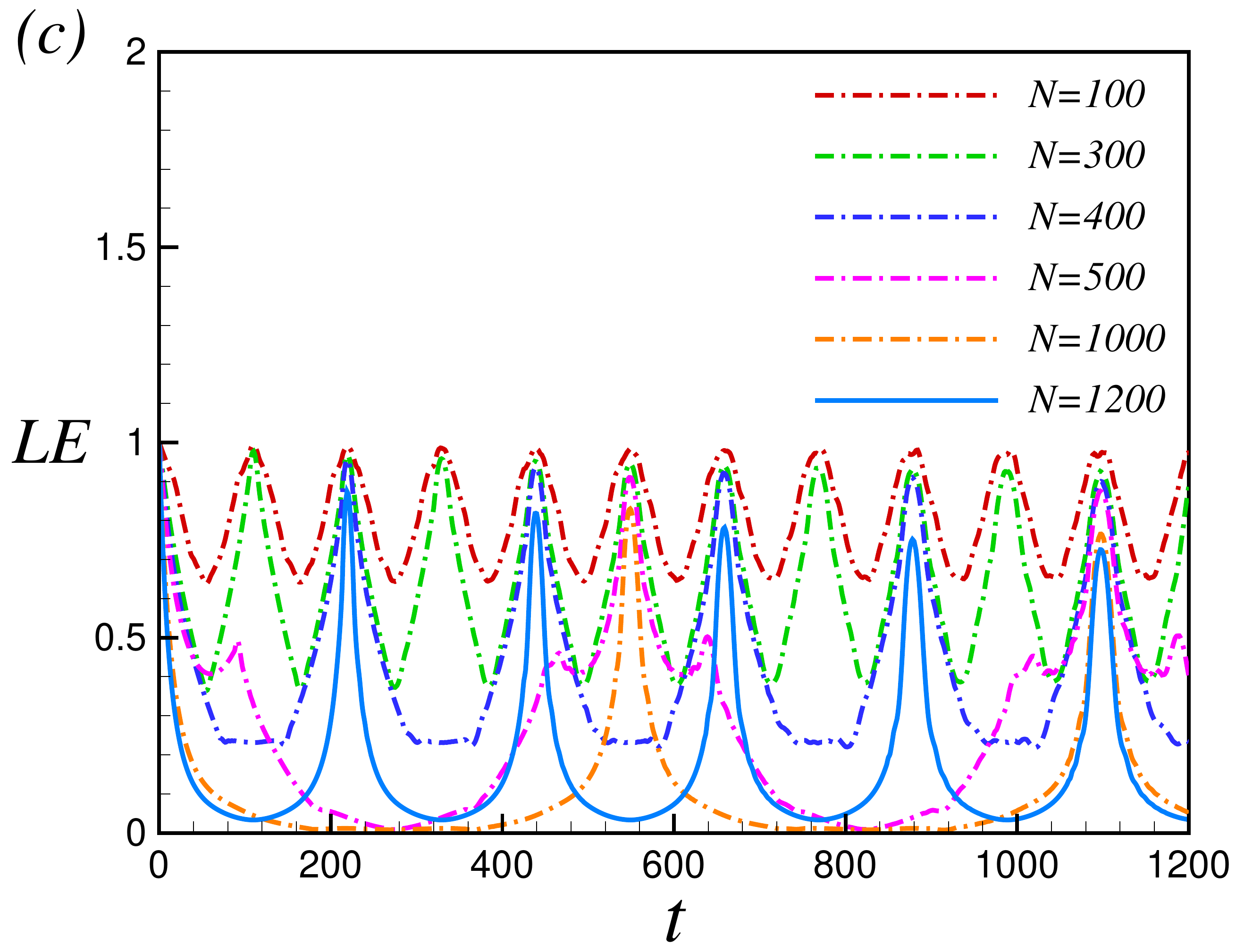}}
\caption{ (Color online) Variation of the LE of the Creutz model versus time $t$ for a quench from $\theta_1=0.0016\pi$ to
the quantum critical point $\theta_c=0$ for different system sizes and for hopping amplitudes (a) $J_{v}=1, J=1$;
(b) $J_{v}=\sqrt{3}, J=1$; and (c) $J_{v}=-1+\sqrt{3}, J=\sqrt{2}$.}
\label{fig5}
\end{figure*}
%

We now show how revivals in the LE can be derived from Eq. (\ref{eq8}). The appearance of a revival requires a large
contribution from all modes in the product of Eq. (\ref{eq8}), equivalent to a small contribution from
the oscillating part of each mode. A numerical analysis shows that the amplitude $A_{k}$ of an oscillation term
is strongly suppressed except close to the critical points $\theta_2\!=\!\theta_{c}\!=\!0, \pi$ in the neighborhoods of the wave numbers $k^{\pm}_c$ of the gap-closing modes
(cf. Fig. \ref{fig3} for the case $\theta_2=0$).
Thus, revivals are controlled by those $k$-modes $\{k^{\pm}_j\}_{j=1,2,...}$ with large oscillation amplitudes $A_{k}$ which cluster around $k^{\pm}_c$:
The first revival time $T_{\text{rev}}$ is the first time instance at which the corresponding oscillating terms vanish. In order to pinpoint $T_{\text{rev}}$, however,  one
must carefully distinguish the case where the gap closes already in the finite system (with $N$ sites on each leg of the ladder) from the case where the gap closing occurs only asymptotically as
$N \rightarrow \infty$. Let us begin by discussing the first case.

When $N$ is finite, the gap closes at the wave numbers $k^{\pm}_c=\pi \pm \arccos(J_{v}/2J)$,  $\Delta\tilde{\varepsilon}_{k}|_{k^{\pm}_c} \!= \!0$, provided that the quantization condition $k^{\pm}_c \!= \!2\pi j^{\pm}_c/N$ is satisfied for some integers $j^{\pm}_c$.
Inspection of Eq. (\ref{eq8}) shows that a revival will appear if the conditions
\begin{equation} \label{cond1}
\Delta\tilde{\varepsilon}_{k}|_{k^{\pm}_{c}-j\delta k}\,t/2 = 0 \, \mbox{mod} \, \pi , \ \ \ j=1,2,...
\end{equation}
are satisfied, with $\delta k=2\pi/N$ the difference between two neighboring modes in the
large-amplitude cluster with wave numbers $\{k^{\pm}_j\}_{j=1,2,...}$.
Using that $\Delta\tilde{\varepsilon}_{k}|_{k^{\pm}_c} = 0$, a first-order Taylor expansion of $\Delta\tilde{\varepsilon}_{k}|_{k^{\pm}_{c}-j \delta k}$ at $k\!=\!k^{\pm}_{c}$, $\Delta\tilde{\varepsilon}_{k}|_{k^{\pm}_{c}-j\delta k}\approx - j\delta k\,\partial \Delta\tilde{\varepsilon}_{k}/\partial k|_{k^{\pm}_{c}}$,
manifests that modes near $k^{\pm}_{c}$ contribute to the revival whenever
$t$ is a multiple of $N/|v_{g}|$ where $v_{g}=\partial \Delta\tilde{\varepsilon}_{k}/\partial k|_{k^{\pm}_{c}}$ (provided that $t$ is not too large, in case higher-order terms in the expansion may add corrections). The group velocity $v_g$ is that of the quasiparticle excitations in the vicinity of $k^{\pm}_c$, and is the same at $k^+_c$ and $k^-_c$ due to the time-reversal invariance at the critical points $\theta_c =0, \pi$. It follows that, on short and intermediate time scales, the revival period $T^{(N)}_{\text{rev}}$ for a Creutz ladder with $N$ sites on each leg is given by
%
\bea
\label{eq10}
T^{(N)}_{\text{rev}}\approx \frac{N}{|v_{g}|}.
\eea
%
To summarize: For a finite system with gap-closing modes $k^{\pm}_c$, periodic revivals occur when oscillation terms with large amplitudes in the mode expansion of the LE, Eq. (\ref{eq8}),
vanish simultaneously with the $k^{\pm}_c$-terms (which are the ones with the largest amplitudes).

In Fig. (4a) the time evolution of the LE following on a quench from $\theta_{1}=0.0016\pi$
to the critical point $\theta_{2}=0$ has been plotted for different system sizes, choosing $J_{v}=J=1$.
For this choice of hopping amplitudes, $|v_{g}|=2\sqrt{3}$ and the gap-closing modes are at $k^{\pm}_{c}=\pi \pm \pi/3$.
An analysis of the data in Fig. (4a) shows that the revivals are governed by Eq. (\ref{eq10}) only when $N$ is divisible by 3 (solid lines in Fig. (4a)).
This confirms our analysis in Sec. II: Eq. (\ref{eq10}) is conditioned on the existence of a gap-closing mode, which in turn is conditioned on
the satisfiability of Eq. (\ref{Ncondition}) with $p=1, q=3$ (given that $k^{\pm}_c = \pi \pm \pi/3$), thus implying that $N$ must be divisible by 3.

This poses the question: How to understand the longer periods of the LE revivals for the systems in Fig. (4a) when $N\neq 3m, m \in \mathbb{N}$ (dashed-dotted lines in the figure)?
For these values of $N$ the system does not contain the gap closing
modes $k^{\pm}_c$ and hence Eq. (\ref{eq10}) does not apply. Still, as seen in Fig. (4a), the revivals are periodic.
What then governs these revivals?
To answer this question, let us go back to Eq. (\ref{eq8}) and recall that periodic revivals occur when its oscillation terms with large amplitudes
vanish simultaneously with the $k^{\pm}_c$ terms. Since now $\Delta\tilde{\varepsilon}_{k}|_{k^{\pm}_c} \!\neq \!0$, Eq. (\ref{cond1}) must be
extended to include also the (would be) gap-closing modes $k_c^{\pm}$ (when $N \rightarrow \infty$):
\begin{equation}  \label{cond2}
\Delta\tilde{\varepsilon}_{k}|_{k_c^{\pm} - j \delta k} \,t/2 = 0 \, \mbox{mod} \, \pi ,  \ \ \ j=0,1,2,...,
\end{equation}
with $\delta k = 2\pi/N$. Considering that none of the modes in Eq. (\ref{cond2}) satisfy the quantization conditions for the given system size and hence are not allowed,
we must instead aim attention at the {\em allowed $k$-modes which are closest to these modes.}

For concreteness, let us consider the case $N=100$, with the LE displayed in Fig. (4a). The nearest large-amplitude oscillation mode to $k^-_c=2\pi/3$ can be written, in obvious
notation,
\begin{align}
\begin{split}
k^{(100)}_{33} &= \frac{2\pi}{100}\times 33 = \frac{2\pi}{300}\times100 - \frac{2\pi}{300} \\
&= k^{(300)}_{99} = k^{-}_c - \delta k^{(300)}.
\end{split}
\end{align}
It is clear from Eq. (\ref{cond1}) that $T^{(300)}_{\text{rev}}$ satisfies
\begin{equation} \label{cond3}
\Delta{\tilde{\varepsilon}}_{k}|_{k_{99}^{(300)}} T^{(300)}_{\text{rev}}/2 =  0 \,\mbox{mod} \,\pi.
\end{equation}
As follows from Eq. (\ref{cond2}) (with $k_c^{-}$ replaced by the nearest allowed mode $k_{33}^{(100)}$), the first revival time $T_{\text{rev}}^{(100)}$ is
obtained from
\begin{equation} \label{cond4}
\Delta{\tilde \varepsilon}_{k}|_{k_{33}^{(100)}-j\delta k^{(100)}}T^{(100)}_{\text{rev}}/2 = 0\, \mbox{mod} \,\pi,
\end{equation}
with $j=0,1,2,...$ as before indexing the wave numbers
in the corresponding large-amplitude cluster.
By inspection, Eqs. (\ref{cond3}) and (\ref{cond4}) are fulfilled simultaneously whenever $T^{(100)}_{\text{rev}}$ is a multiple of $T^{(300)}_{\text{rev}}$, where, according to
(\ref{eq10}),
\begin{equation}
T^{(300)}_{\text{rev}}\approx \frac{300}{|v_{g}|},
\end{equation}
with $|v_g|=2\sqrt{3}$. Generalizing to an arbitrary finite system with $N \neq 3m, m \in \mathbb{N}$,
it follows that the revival period of the LE for a Creutz ladder with $J_v=J=1$
is given by the same expression as in Eq. (\ref{eq10}) but with $N$ replaced by
$N^{'}$, the least common multiple (LCM) of 3 and $N$,

\bea
\label{eq13}
T^{(N\neq3m)}_{\text{rev}}\approx \frac{N^{'}}{|v_{g}|}, \ N^{'}=\mbox{LCM}(3,N), \ m \in \mathbb{Z}^+.
\eea
Table \ref{table1} displays the revival periods for different system sizes obtained from Eq. (\ref{eq13}) (valid for short and intermediate time scales),
showing excellent agreement with the numerical data in Fig. (4a).

%
\begin{table}[h!]
\centering
\caption{First revival time for different system sizes when $J_{v}=J=1$.
The values of $N$ correspond to the dashed-dotted lines in Fig. (4a).}
\begin{tabular}{ |c || c||c| }
\hline
 $N$ & $T^{(N)}_{\text{rev}}$ & $T^{(Simulation)}_{\text{rev}}$\\
 \hline\hline
  $100$  &  $T^{(300)}_{\text{rev}}=300/v_{g}=300/2\sqrt{3}\simeq86.60$ & $86.58$   \\
 \hline
 $400$  &  $4T^{(300)}_{\text{rev}}=1200/v_{g}=1200/2\sqrt{3}\simeq346.41$  &  $345.89$ \\
 \hline
 $500$ &  $5T^{(300)}_{\text{rev}}=1500/v_{g}=1500/2\sqrt{3}\simeq433.0$ & $432.33$ \\
 \hline
 $1000$ &  $10T^{(300)}_{\text{rev}}=3000/v_{g}=3000/2\sqrt{3}\simeq866.0$ &  $865.44$ \\
  \hline
\end{tabular}
\label{table1}
\end{table}
%
%
\begin{table}[h!]
\centering
\caption{First revival time for different system sizes when $J_{v}=\sqrt{3}$, $J=1$.
The values of $N$ correspond to the dashed-dotted lines in Fig. (4b).}
\begin{tabular}{ |c || c||c| }
\hline
 $N$ & $T^{(N)}_{\text{rev}}$ & $T^{(Simulation)}_{\text{rev}}$\\
 \hline\hline
 $100$  &  $T^{(300)}_{\text{rev}}=300/v_{g}=300/2\simeq150.00$ &  $149.22$\\
 \hline
 $400$ &  $4T^{(300)}_{\text{rev}}=1200/v_{g}=1200/2\simeq600.00$ & $599.69$ \\
 \hline
 $500$ &  $5T^{(300)}_{\text{rev}}=1500/v_{g}=1500/2\simeq750.00$  & $751.00$ \\
  \hline
 $1000$ &  $10T^{(300)}_{\text{rev}}=3000/v_{g}=3000/2\simeq1500.00$ & $1499.97 $ \\
 \hline
\end{tabular}
\label{table3}
\end{table}
%

%
\begin{table}[h!]
\centering
\caption{First revival time for different system sizes when $J_{v}=-1 + \sqrt{3}$, $J=1$.
The values of $N$ correspond to the dashed-dotted lines in Fig. (4c).}
\begin{tabular}{ |c || c||c| }
\hline
 $N$ & $T^{(N)}_{\text{rev}}$ & $T^{(Simulation)}_{\text{rev}}$\\
 \hline\hline
 $100$  &  $T^{(600)}_{\text{rev}}=T^{(1200)}_{\text{rev}}/2=600/v_{g}\simeq109.80$ & $109.39$ \\
 \hline
 $300$  &  $T^{(600)}_{\text{rev}}=T^{(1200)}_{\text{rev}}/2=600/v_{g}\simeq109.80$ &  $109.70$\\
 \hline
 $400$ &  $2T^{(600)}_{\text{rev}}=T^{(1200)}_{\text{rev}}=1200/v_{g}\simeq219.61$ & $219.27$ \\
 \hline
 $500$ &  $5T^{(600)}_{\text{rev}}=5T^{(1200)}_{\text{rev}}/2=3000/v_{g}\simeq549.03$ & $549.38$ \\
 \hline
 $1000$ &  $5T^{(600)}_{\text{rev}}=5T^{(1200)}_{\text{rev}}/2=3000/v_{g}\simeq549.03$ & $549.30$  \\
 \hline
\end{tabular}
\label{table4}
\end{table}
%
Let us examine two additional cases. Figs. (4b) and  (4c) exhibit numerical data for the same quench as before, from $\theta_{1}=0.0016\pi$ to the critical point $\theta_{2}=0$,
but now for the Creutz ladder with hopping amplitudes $J_{v}=\sqrt{3},J=1$ and $J_{v}=-1+\sqrt{3},J=\sqrt{2}$, respectively.  In the first case, carrying out the same analysis as above, the revival period of the LE when $N=12m\ (m=1,2,...)$ is predicted to be given by Eq. (\ref{eq10}) with $|v_g|=2$, while for $N\neq 12m, \, T^{(N\neq12m)}_{\text{rev}}=N^{'}/|v_{g}|,  \ N^{'}=\mbox{LCM}(12,N).$ Similarly, in the second case, where $v_{g} = 2\sqrt{4+2\sqrt{3}}$, the revival period is again predicted to be given by Eq. (\ref{eq10}) when $N=24m \ (m=1,2,...)$,
now with $|v_{g}| = 2\sqrt{2(2+\sqrt{3})}$, while for $N\neq 24m, \, T^{(N\neq24m)}_{\text{rev}}=N^{'}/|v_{g}|, \ N^{'}=\mbox{LCM}(24,N).$
Again, the agreement with the numerical data (solid [dashed-dotted] lines in Figs. \ref{fig5}(b) and  \ref{fig5}(c) for
commensurate [incommensurate] system sizes) is excellent, as can be read off from the tabulated incommensurate
revival periods in Tables \ref{table3} ($N \neq 12m$) and \ref{table4} ($N \neq 24m$).
Yet other choices of hopping amplitudes $-$ implying other zero-energy modes and hence other commensurability
conditions for the system size $-$ produce equally satisfying agreement between theory and numerical data.

In summary, the revivals of the LE for finite Creutz ladders after a quench to one of the equilibrium quantum critical points do not scale
linearly with the size of the system, contrary to what has been found for the post-quench LE of other models \cite{Montes,Happola,Cardy,Michal}.
It has been shown that the revivals are controlled
by the modes in the neighborhood of the gap-closing zero-energy modes for which the oscillation amplitudes in the mode decomposition of the LE are the largest.
Since information propagates through the system via the wave packets of quasiparticles,
the revival times can thus be identified as the time instances at which quasiparticles associated with large LE oscillation amplitudes
are synchronized with the zero-energy modes. We would here like to direct attention to a result by H\"app\"ol\"a {\em et al.} \cite{Happola}, showing that the odd-numbered revivals in the LE
of the transverse field Ising chain subject to antiperiodic boundary conditions do not appear when instead periodic boundary conditions are used. In other words, the periodicity of revivals when using
periodic boundary conditions
is twice that for the case when antiperiodic boundary conditions are imposed. This feature may be explained by our finding in this paper. It is straightforward to show that in the case of antiperiodic boundary conditions
(which is the proper boundary condition to impose when analyzing the model with an even number of sites using fermionization, as done in Ref. \cite{Happola}), the system contains a zero-energy mode at $k=0$, while for periodic boundary conditions (to be used when the number of sites is odd) there is no zero-energy mode. Thus, carrying over our result for the Creutz model to the transverse field Ising model mapped onto a fermionic model as in Ref. \cite{Happola}, the expression for the revival time $T^{(N)}_{rev}$ in the case of periodic boundary conditions (odd number of sites $N$) is seen to be given by $T^{N}_{rev}=\mbox{LCM}(2,N)/|v_g|$, with $v_g$ the quasiparticle group velocity in the neighborhood of $k=0$. This explains the period-doubling compared to the case of antiperiodic boundary conditions.
It is worthwhile to mention that, the connection between dynamic finite-size scaling and critical exponents has been recently studied in Ref. \cite{Pelissetto} for both continuous and first-order quantum transitions, which could be interesting to be applied on Creutz model in future studies.
%
%
\section{Dynamical Quantum Phase Transitions \label{DQPT}}
As discussed in the Introduction, there has recently been a growing interest in the study of dynamical phase transitions (DQPTs),
probing non-analyticities in the complex time plane of the dynamical free energy density \cite{Fagotti} of a quenched system \cite{Zvyagin2016,HeylReview}.
An early result for a DQPT following a sudden quantum quench in the one-dimensional transverse field Ising model,
reported by Heyl {\em et al.} \cite{Heyl2013}, suggested that DQPTs occur only if
the quench is performed across an equilibrium quantum critical point.
Further studies, however, revealed that DQPTs can occur following a quench also within the same phase
\cite{Andraschko,Sharma2015,Vajna}. Other theoretical works have explored DQPTs in topological and mixed phases \cite{Vajna2015,Budich2016,Budich2017,Bhattacharya2017,Bhattacharya2017b}
and also after slow quenches (``ramps") \cite{Sharma2016,Divakaran2016,Tatjana}.

The concept of a DQPT draws on the similarity between the canonical partition function
of an equilibrium system $Z(\beta)=\mbox{Tr} e^{-\beta H}$ and the boundary quantum partition function
$Z(z)=\langle\psi_{0}|e^{-z H}|\psi_{0}\rangle$ with $|\psi_0\rangle$ a boundary state and $z \in \mathbb{C}$ \cite{LeClair, Piroli}. When $z=it$,
the boundary quantum partition
function becomes equivalent to a Loschmidt amplitude (LA), $L(t)=\langle\psi_{0}|e^{-i Ht}|\psi_{0}\rangle$,
the modulus of which defines a Loschmidt echo ${\cal L}(t)$ (cf. Eq. (\ref{eq7})).
Using our notation for the Creutz ladder, the LA  given by $L(t)=\langle\Psi_{G}(\theta_1)|e^{-i H(\theta_2)t}|\Psi_{G}(\theta_1)\rangle$
is the overlap amplitude of the initial quantum state $|\Psi_{G}(\theta_1)\rangle$ with its time-evolved state controlled by the post-quench
Hamiltonian $ H(\theta_2)$, and where the ground state $|\Psi_{G}(\theta_1)\rangle$ stands in for a boundary state. Defining the free energy density $f(z)$ in the complex time plane as
$f(z)=-\lim_{N\rightarrow\infty}\ln Z(z)/N$, with $N$ the number of degrees of freedom, $f(z)\  [Z(z)]$ is frequently referred to as the {\em dynamical} free energy density
[partition function] \cite{Fagotti,HeylReview}.
In the spirit of classical equilibrium statistical mechanics \cite{YangLee,LeeYang}, one then searches for non-analyticities in $f(z)$, or zeros of $Z(z)$
(known as Fisher zeros \cite{Fisher1978}), now interpreted as signals of DQPTs \cite{Heyl2013,HeylReview}.
Additionally, these DQPTs are imprinted as nonanalyticities in the rate function $l(t)$ of the Loschmidt echo  \cite{Pollmann,Heyl2013,Andraschko, Sharma2015},
with $l(t)$ defined as
\begin{equation} \label{rate}
l(t)=-\lim_{N\rightarrow\infty}\ln{\cal L}(t)/N.
\end{equation}

\begin{figure*}
\centerline{\includegraphics[width=0.33\linewidth]{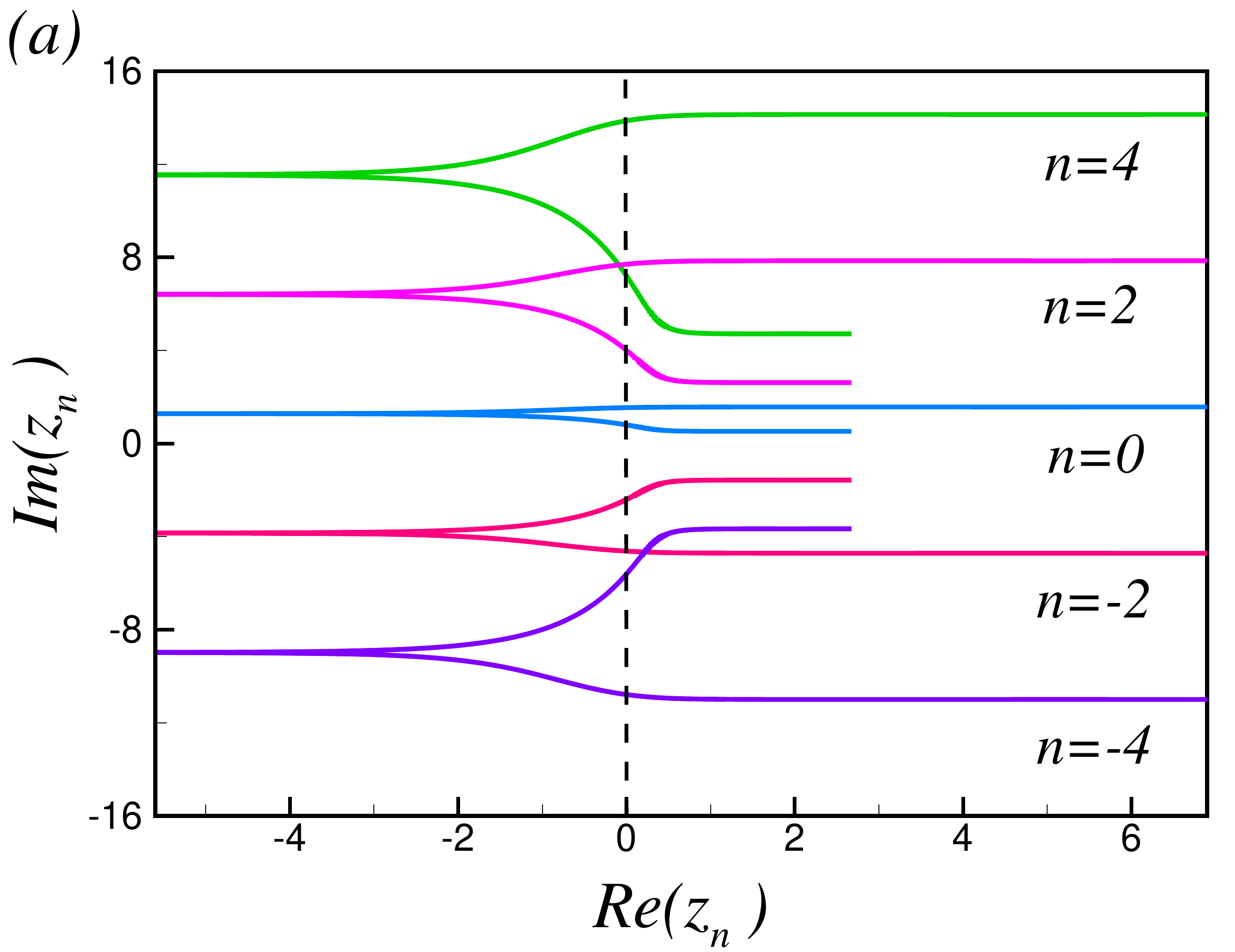}
\includegraphics[width=0.33\linewidth]{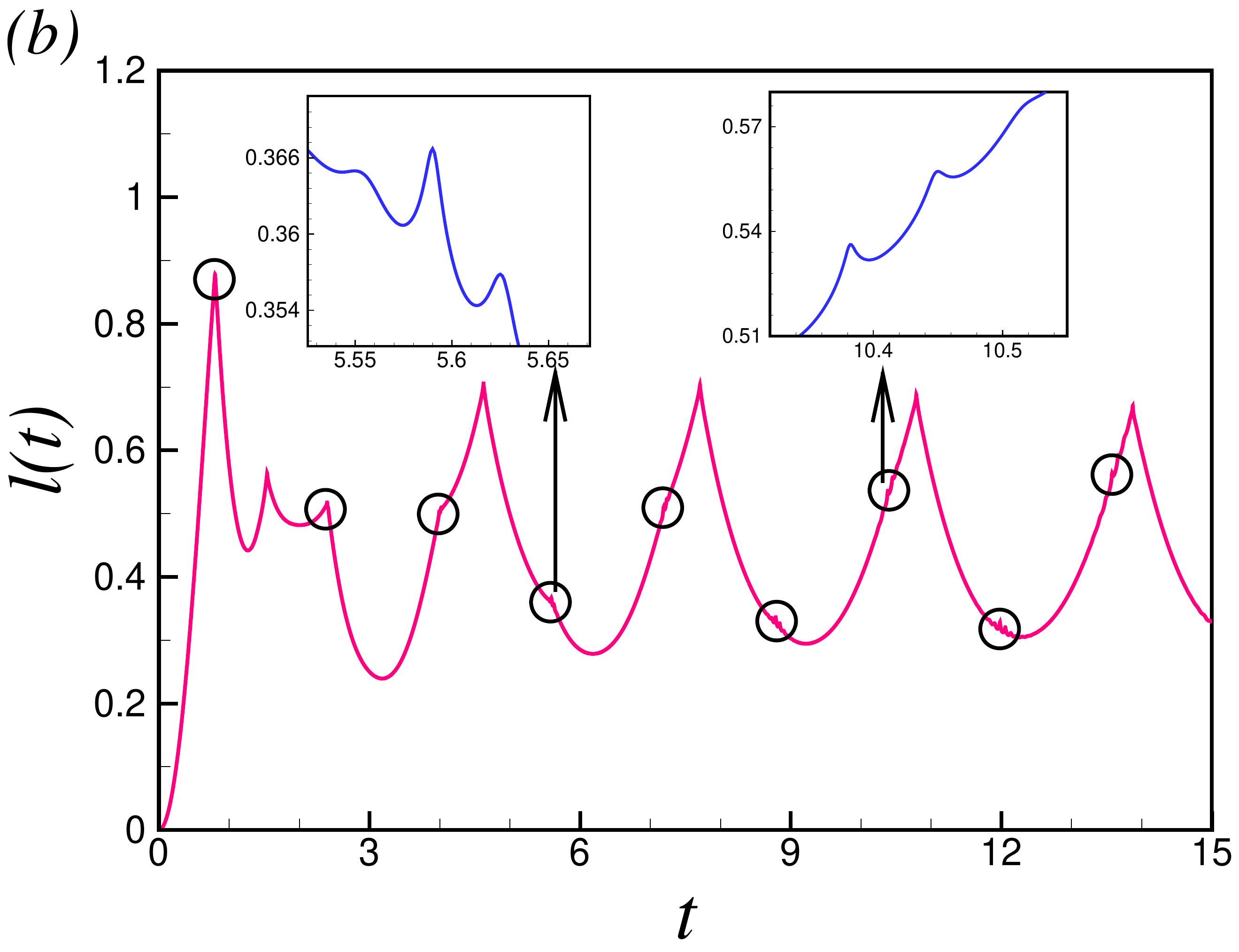}
\includegraphics[width=0.33\linewidth]{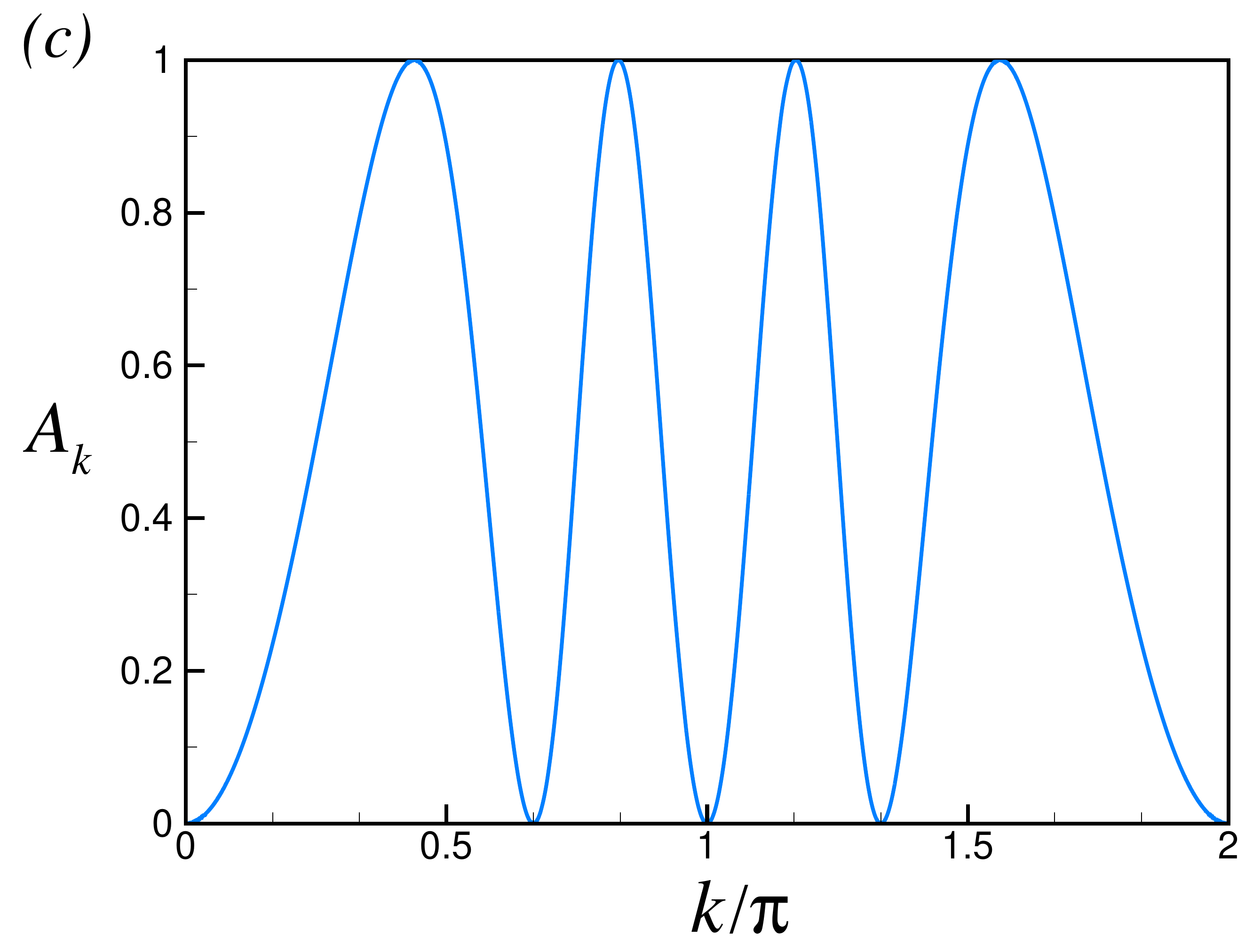}}
\caption{ (Color online) (a) Lines of Fisher zeros for a quench across the equilibrium quantum
critical point $\theta_c=0$ from $\theta_{1}=0.25\pi$ to $\theta_{2}=-0.25\pi$.
The different lines correspond to different values of $n$ in Eq. (\ref{eq18}); (b) Cusps in the return rate $l(t)$, defined in Eq. (\ref{rate})
indicate DQPTs after the quench from $\theta_{1}=0.25\pi$ to $\theta_{2}=-0.25\pi$.
Insets zoom in on shorter nonequilibrium timescales. (c) The amplitude $A_{k}$ in Eq. (\ref{eq9}) versus wave number $k$ for the quench from
$\theta_{1}=0.25\pi$ to $\theta_{2}=-0.25\pi$. In all figures (a), (b), and (c), $J=J_v=1$, and $N=9000$.}
\label{fig6}
\end{figure*}

It is straightforward to show that the dynamical partition function corresponding to the ground state of the Creutz model is given by
%
\bea
\label{eq17}
Z(z)\!=\!\prod_{k}e^{-z\tilde{\varepsilon}^{\alpha}_{k}(\theta_{2})}\!\big(\!\cos^{2}(\eta_{k})\!+\!\sin^{2}(\eta_{k})e^{-z\Delta\tilde{\varepsilon}_{k}(\theta_{2})}\big)
\eea
%
with $\tilde{\epsilon}^{\alpha}_k$ defined after Eq. (\ref{eq6}), and $\eta_k$ and $\Delta\tilde{\epsilon}_k$ defined in Eq. (\ref{eq9}).
The zeros of the LA in the complex plane form a family of lines
labeled by an integer $n$,
%
\bea
\label{eq18}
z_{n}(k)= \frac{1}{\Delta\tilde{\varepsilon}_{k}}\Big[i\pi(2n+1)+\ln(\tan^{2}(\eta_{k}))\Big].
\eea
%
A plot of lines of Fisher zeros is depicted in Fig. \ref{fig6}(a) for a quench from
$\theta_{1}=0.25\pi$ to $\theta_{2}=-0.25\pi$. This quench is performed across the equilibrium quantum critical point $\theta_c=0$,
with the lines of Fisher zeros crossing the imaginary axis in the complex time plane.
%
%
The main quantity that controls the dynamical free energy is $\tan^{2}(\eta_{k})$, which depends on the
parameters of the initial ("pre-quench") and final ("post-quench") Hamiltonian (with the initial state being the ground state of the
pre-quench Hamiltonian).
As seen from Eq. (\ref{eq18}), the lines of Fisher zeros cross the imaginary axis only when there is a mode $k^{\ast}$
that satisfies $\tan^{2}(\eta_{k^{\ast}})=1$. Using the expression
$2\eta_{k}=\gamma_{k}(\theta_{1})-\gamma_{k}(\theta_{2})$ and Eq. (\ref{eq4}), this condition
can be rewritten as
%
\bea
\label{eq19}
(2J\!\cos(k^{\ast})\!+\!J_{v})^{2}\!=\!-(4J\sin(k^{\ast}\!))^{2}\!\sin(\theta_{1})\sin(\theta_{2}).
\eea
%
This equation can be fulfilled only when $\sin(\theta_{1})\sin(\theta_{2})$ is negative semidefinite.
In other words, the non-analyticities in the LA exist only when the quench is performed
{\em across} one of the critical points $\theta_c=0, \pi$ or {\em to} $\theta_c=0$ or $\theta_c=\pi$.
Given Eq. (\ref{eq18}) with $z_n=it$, it follows that the rate function $l(t)$ of the LE
shows a periodic sequence of real-time nonanalyticities for quenches across or to one of the critical points $\theta_c=0, \pi$
at times
%
\bea
\label{eq20}
t_{n}=t^{\ast}(n+\frac{1}{2}),~~t^{\ast}=\frac{2\pi}{\Delta\tilde{\varepsilon}_{k^{\ast}}}.
\eea
%
This result is in agreement with the numerical data shown in Fig. \ref{fig6}(b), obtained for a quench from $\theta_{1}=0.25\pi$ to $\theta_{2}=-0.25\pi$.
 Cusps in $l(t)$ are clearly visible as signs of DQPTs.
It is important to note that as the imaginary axis is crossed twice by the lines of Fisher zeros there are two timescales in the dynamical free
energy. The cusps marked by circles in Fig. \ref{fig6}(b) correspond to the shorter nonequilibrium scale.
%
\begin{figure*}[t!]
\centerline{\includegraphics[width=0.33\linewidth]{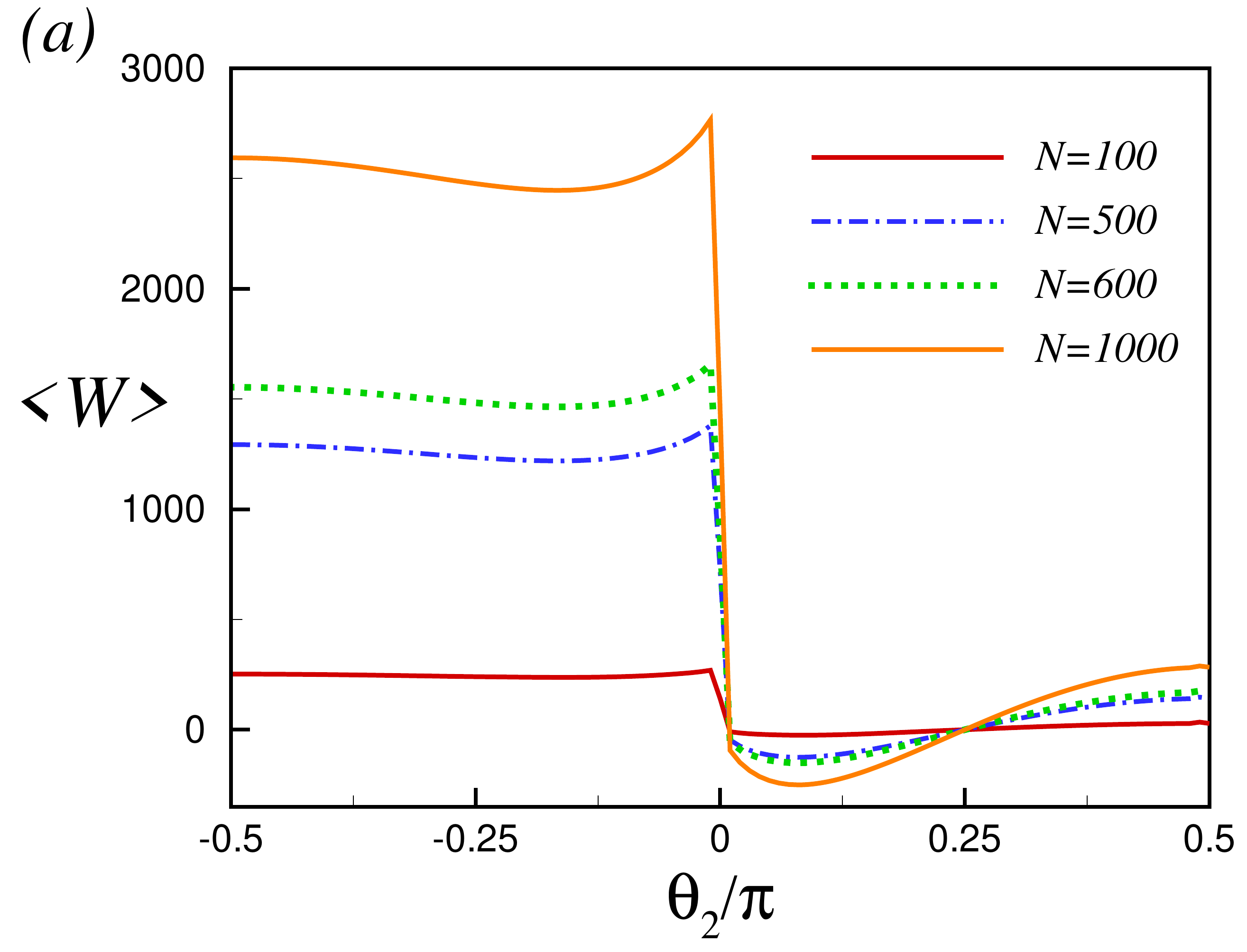}
\includegraphics[width=0.33\linewidth]{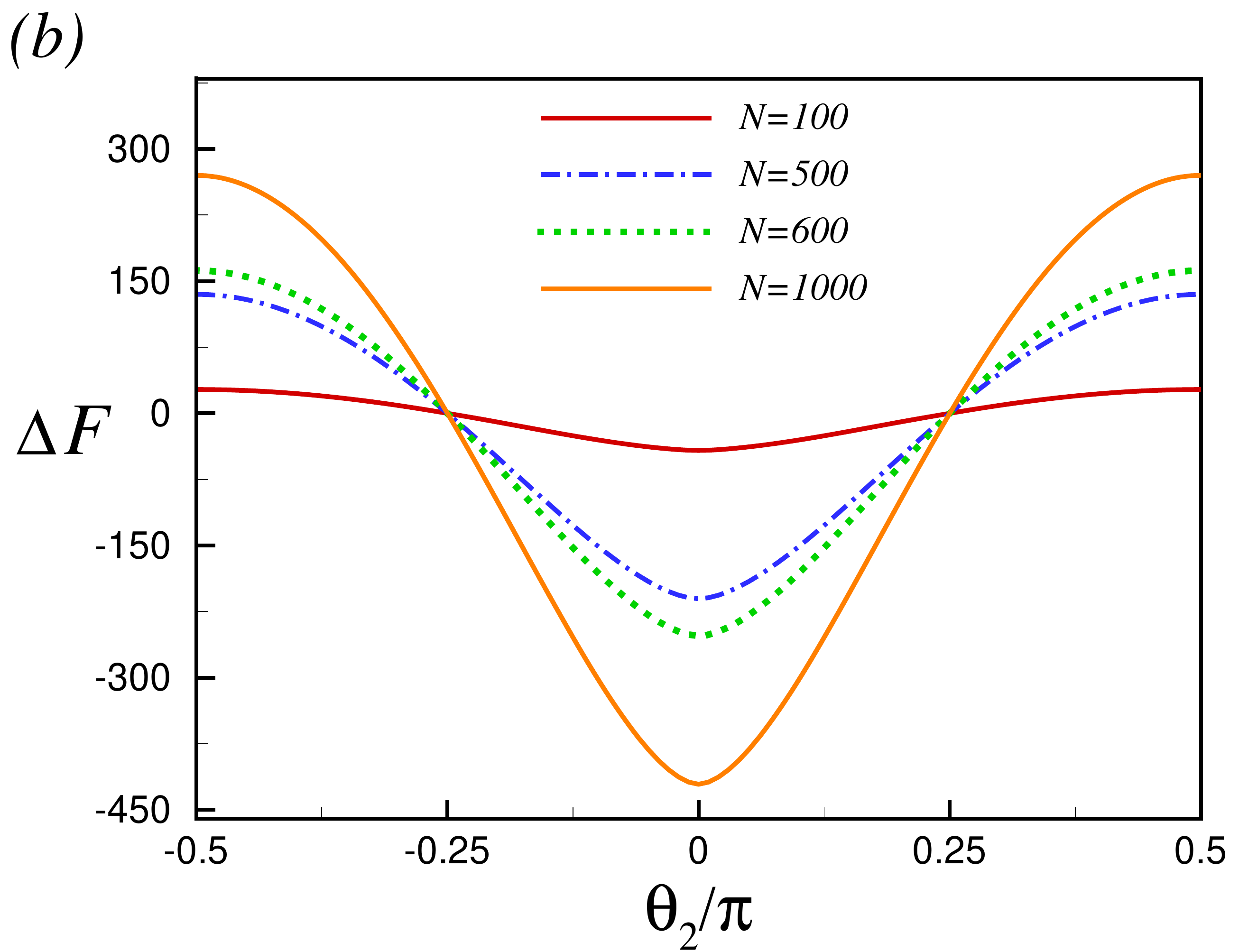}
\includegraphics[width=0.33\linewidth]{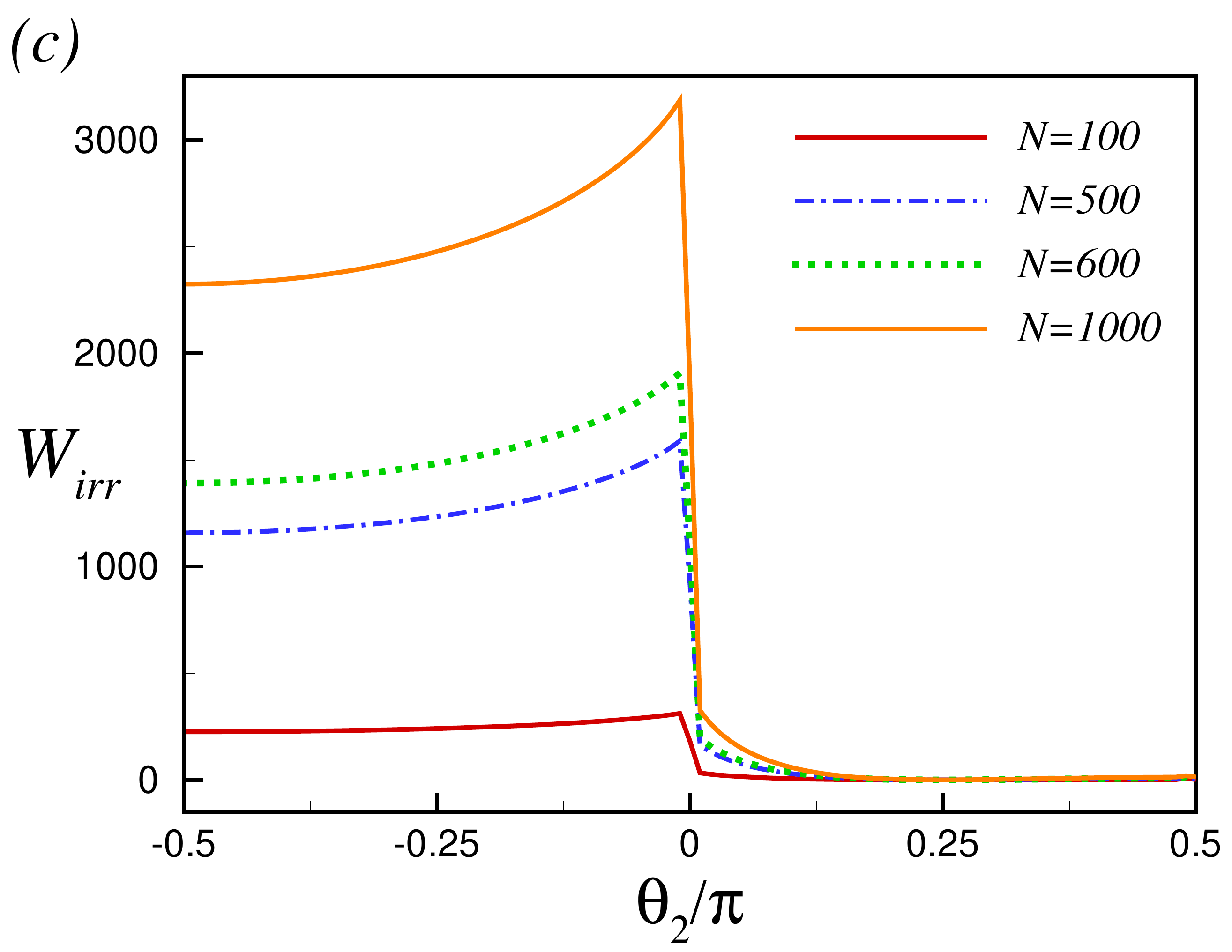}}
\caption{ (Color online) (a) Average work $\langle W \rangle$, (b) difference of ground-state energies $\Delta F=E_g(\theta_2) - E_g(\theta_1)$, and (c) irreversible
work $W_{\text{irr}}$ versus $\theta_{2}/\pi$ for a quench from $\theta_{1}=0.25\pi$ to $\theta_{2}$
for different system sizes when $J_{v}=J=1$.}
\label{fig7}
\end{figure*}
%

To better understand the origin of the nonanalyticities in $l(t)$, let us take a closer look at
the LE in Eq. (\ref{eq8}). First recall that the real-time nonanalyticities coincide with
the time instances at which the LE vanishes \cite{Heyl2013,HeylReview}. This happens only if one factor in the mode decomposition in Eq. (\ref{eq8}) becomes zero,
concurrent with the oscillating part of the $k^{\ast}$-mode becoming equal to unity. An analysis shows that the oscillation
amplitude $A_{k}$ is small for a quench within the same phase, while it takes its maximum possible value
($A_{k}=1$) at $k=k_{1}^{\ast}$ and $k=k_{2}^{\ast}$ when a quench crosses $\theta_{c}=0$ (Fig. \ref{fig6}(c)).
Consequently, the corresponding $k_{1,2}^{\ast}$ modes can contribute destructively to the LE only at time
instances for which the associated oscillating term is unity, i.e.
$A_{k_{1,2}^{\ast}}\sin^{2}(\Delta\varepsilon_{k_{1,2}^{\ast}}t/2)=1$.
This requires that $t_{n}=(2n+1)\pi/\Delta\varepsilon_{k_{1,2}^{\ast}}=t_{1,2}^{\ast}(n+1/2)$,
which is exactly the same condition as in Eq. (\ref{eq20}).
Let us mention that the number of modes where the oscillation amplitude takes its maximum possible value ($A_{k}=1$)
can be shown to be equal to the number of time scales in the DQPT \cite{privateJJLM}.

To highlight the important role of zero-energy modes also in DQPTs, let us consider a DQPT in the Creutz ladder when quenching {\em to} (not across) the critical point $\theta_c\!=\!0$.
In such a case, Eq. (\ref{eq19}) is reduced to $2J\cos(k)+J_{v}\!=\!0$, which is fulfilled only for system sizes that contain zero-energy modes (cf. Eq. (\ref{eq9}) with $\epsilon^{qp}_k \!=\! 2J \cos(k) + J_v$ and $\theta_2\!=\!0$).
Thus, while a DQPT can occur in a finite-size Creutz ladder under various circumstances, its appearance after a quench to one of the critical points is conditioned on the presence of zero-energy modes, possible only if the system size is commensurate with the condition in Eq. (\ref{Ncondition}). We expect that this conclusion applies quite generally.
%
%
\section{Magnetic flux quench and work statistics \label{Work}}
The nonequilibrium dynamics of a quenched quantum system can be expressed
in many different ways, borrowing ideas from equilibrium statistical mechanics.
However, since a quench protocol takes the system out of equilibrium, thermodynamic quantities
get replaced by stochastic variables. A case in point is the work $W$ performed by the quench, with $W$
now described by a probability distribution function \cite{Campisi2011},
\begin{eqnarray}\label{wPDF}
p(W)=\sum_{m} | \langle E^{\prime}_m| E_0 \rangle |^2 \delta\left[W\!-\!(E^{\prime}_m{}\!-\!E_0)\right].
\end{eqnarray}
Here $|E_0\rangle$ [$|E^{\prime}_m\rangle$] with corresponding energy $E_0$ [$E^{\prime}_m$] is the ground state
[$m$:th eigenstate] of the pre-quench [post-quench] Hamiltonian. The work probability distribution function in (\ref{wPDF}) is an experimentally
accessible quantity \cite{Dorner2013,Mazzola2013} from which the {\em average work} is obtained as
\begin{equation}
\langle W \rangle = \int W p(W) dW.
\end{equation}

Given the average work $\langle W \rangle$, the Jarzynski fluctuation-dissipation relation \cite{Jarzynski} makes it possible to
define the so called {\em irreversible work}
\begin{equation} \label{Wirr}
W_{\rm irr} = \langle W \rangle -\Delta F \geq 0,
\end{equation}
where $\Delta F$ is the difference between the free energies after and before the quench. At zero temperature (which we assume here),
$\Delta F$ reduces to the difference between the ground-state energies of the post- and pre-quench Hamiltonians: $\Delta F = \Delta E_g=E_g(\theta_2) - E_g(\theta_1)$.
 The irreversible work quantifies the amount
of energy which has to be taken out from the quenched system so that it relaxes to its new equilibrium state $-$ at zero temperature, the ground
state of the post-quench Hamiltonian.

Case studies \cite{Silva,Bayat,Campbell} suggest that the irreversible work $W_{\text{irr}}$ of a quenched system serves as a marker of equilibrium
quantum phase transitions. Here we explore this notion when the equilibrium phase transition is {\em topological}, using the Creutz model as a test case.

Let us begin by writing down a general formula for the average work after a sudden quench \cite{Silva} using our previously defined notation
for the Creutz model,
\bea
\label{eq21}
\langle W \rangle \!&=&\!\langle\Psi_G(\theta_1)|H(\theta_2)|\Psi_G (\theta_{1})\rangle-E_g(\theta_1),
\eea
with $E_g(\theta_1)$ the ground-state energy of the initial Hamiltonian. It is straightforward to translate this into an explicit expression,
{\small
\bea
\label{eq22}
\langle W \rangle\! =\!\sum_{k}\!\Big(\tilde{\varepsilon}^{\alpha}_{k}(\theta_{2})\cos^{2}(\eta_{k})\!+\!\tilde{\varepsilon}^{\beta}_{k}(\theta_{2})\sin^{2}(\eta_{k})
\!-\!\tilde{\varepsilon}^{\alpha}_{k}(\theta_{1})\!\Big),
\eea
}
assuming as before that the Fermi level is at zero energy. By combining Eqs. (\ref{Wirr}), (\ref{eq21}), and (\ref{eq22}), it follows that the irreversible
work after a quench is given by
{\small
\bea
\label{eq23}
W_{\rm irr}\!=\!\sum_{k}\!\Big(\tilde{\varepsilon}^{\alpha}_{k}(\theta_{2})\cos^{2}(\eta_{k})\!+\!\tilde{\varepsilon}^{\beta}_{k}(\theta_{2})\sin^{2}(\eta_{k})
\!-\!\tilde{{\varepsilon}}^{\beta}_{k}(\theta_{2})\!\Big).
\eea
}
%
\begin{figure*}[t!]
\centerline{\includegraphics[width=0.33\linewidth]{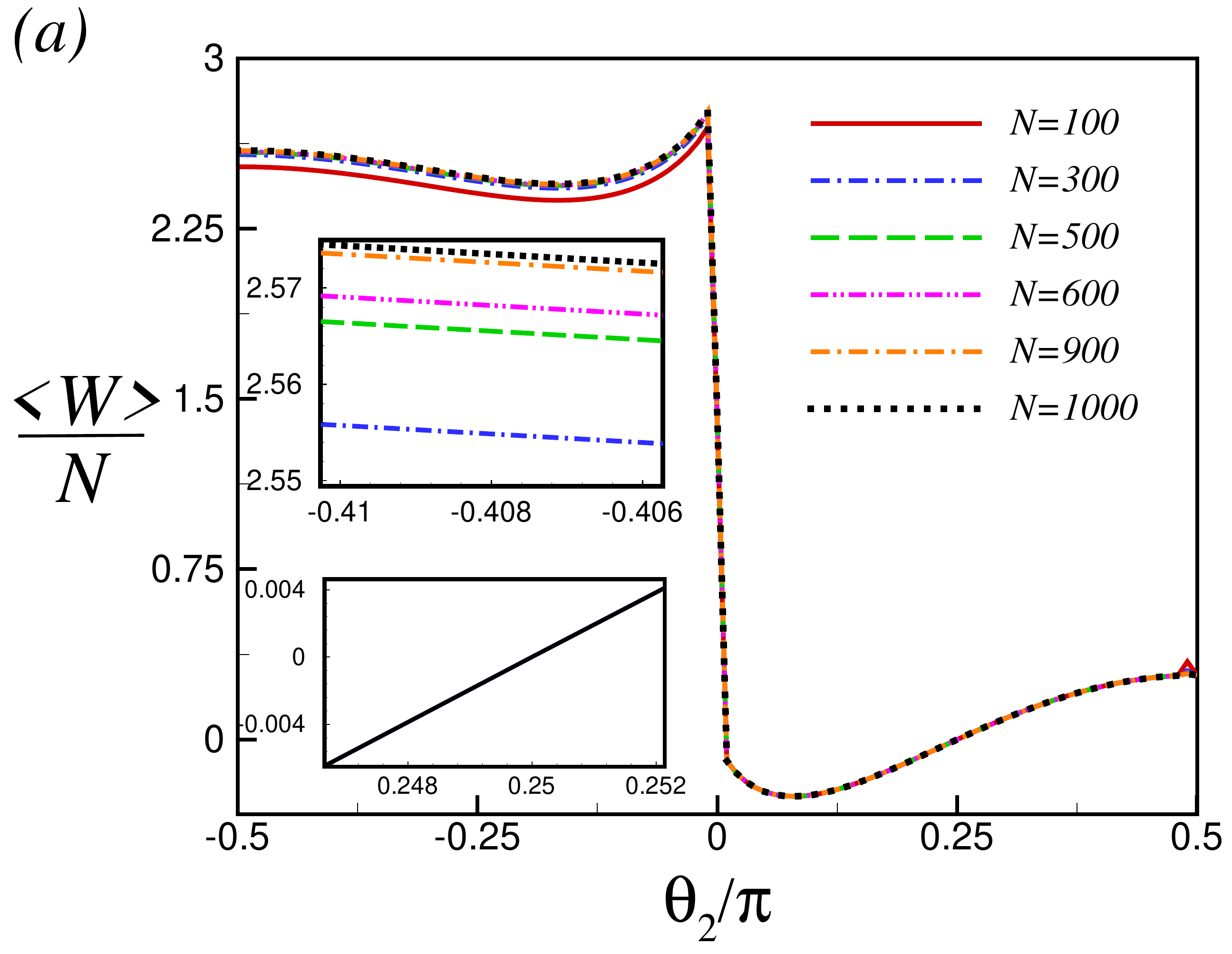}
\includegraphics[width=0.33\linewidth]{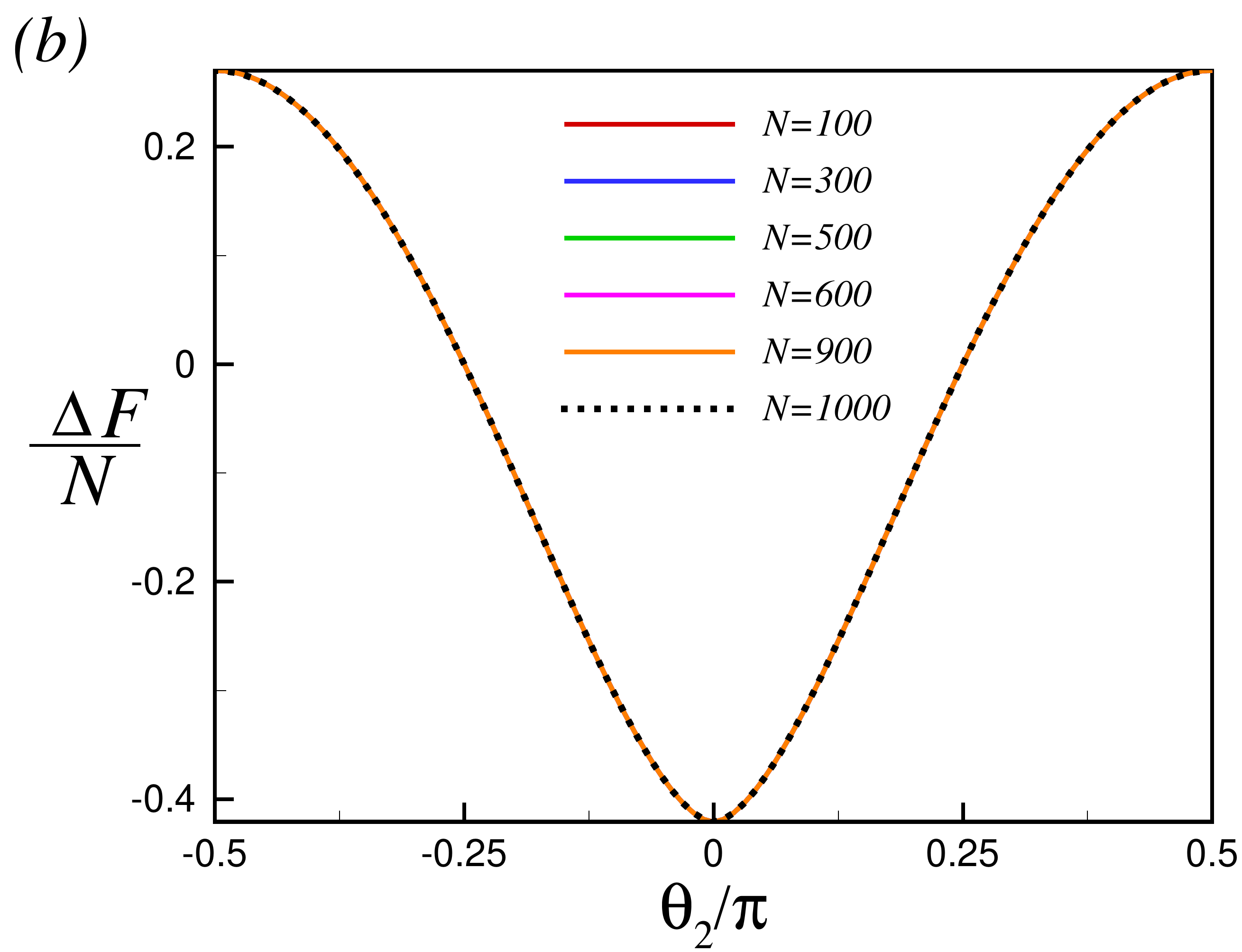}
\includegraphics[width=0.33\linewidth]{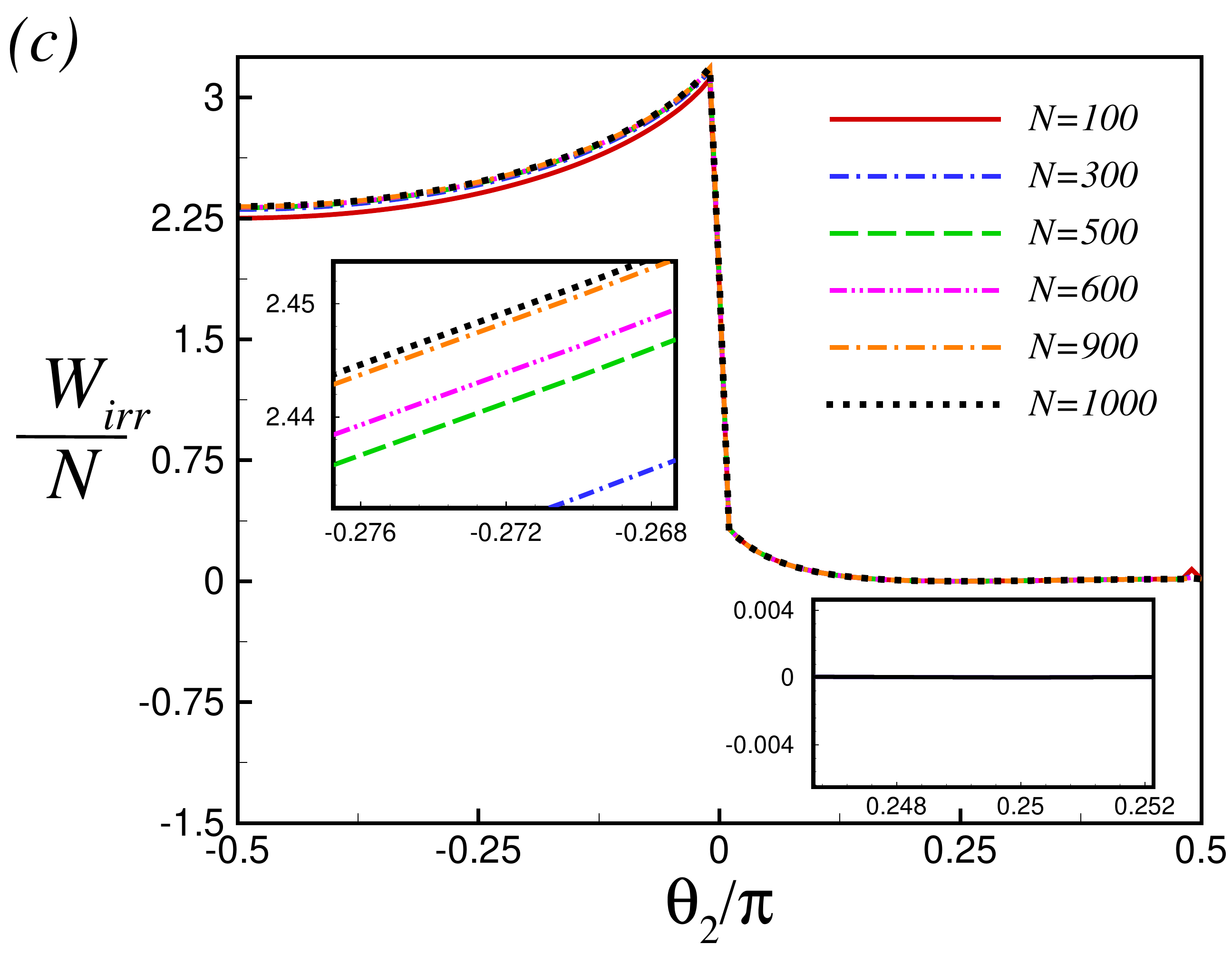}}
\caption{ (Color online) (a) Average work per particle $\langle W \rangle/N$, (b) change of ground-state energy per particle $\Delta F/N$, and
(c) irreversible work per particle $W_{\rm irr}/N$ versus $\theta_{2}$ for a quench from
$\theta_{1}=0.25\pi$ to $\theta_{2}$ for different system sizes when $J_{v}=J=1$.
The insets zoom in on the fine structure of the corresponding curves.}
\label{fig8}
\end{figure*}
%
In Fig. \ref{fig7} the average work $\langle W \rangle$, the change of the ground-state energy $\Delta F$, and the irreversible work $W_{\rm irr}$ have been plotted against $\theta_{2}$ for
different system sizes, for a quench from fixed $\theta_{1}=0.25\pi$ to $\theta_{2}$.
Recalling that $\theta_c=0$ is a quantum critical point, the numerical data in Fig. \ref{fig7}(a) show that $\langle W \rangle $ is overall small for a quench within
the same phase.
Positive [negative] values of $\langle W \rangle $ reveal a quench by which the magnetic flux is increased [decreased].
Thus, as expected, $\langle W \rangle=0 $ corresponds to the case of no quench at all. For a quench crossing the critical point $\theta_2 \!= \!\theta_c \!=\! 0$,
$\langle W \rangle $ takes positive and large values and increases with the system size.

As seen in Fig. \ref{fig7}(b), the change of the ground-state energy of the post-quench Hamiltonian is symmetric with respect to the critical point $\theta_2=0$
where it takes its minimum. The change of the ground-state energy is positive when quenching the system to a point where $|\theta_{2}|$
is larger than $|\theta_{1}|$. In contrast, it becomes negative for quenching the magnetic flux to values smaller than $|\theta_{1}|$.

Fig. \ref{fig7}(c) shows that when the quench is confined to the same phase as the initial state,
the irreversible work vanishes away from the critical point, indicating that the process is fully reversible.
This is to be compared to a quench into the neighborhood of the critical point within the same phase where $W_{\rm irr}$  takes small nonzero values.
Differently, the irreversible work becomes quite large when the quench crosses the critical point, making manifest the irreversibility
of the process, with  $W_{\rm irr}$ increasing with system size.

The average work per particle $\langle W \rangle/N$, change of ground-state energy per particle $\Delta F/N$, and irreversible work per particle $W_{\rm irr}/N$ are depicted in
Fig. \ref{fig8}. As is evident from figures \ref{fig8}(a) and (c) (cf. bottom inset), $\langle W \rangle /N$ and $W_{\rm irr}/N$ are independent of system size for quenches
within the same phase. In other words, for these cases $\langle W \rangle$ and $\Delta F$ are clearly extensive, as expected for a noninteracting system.
Remarkably, when quenching the system {\em through} the quantum critical point $\theta_c=0$, curves for different system sizes do not exhibit perfect data collapse.
Although the violation is small and visible only in the fine structure of the curves (top insets in Figs. \ref{fig8}(a) and (c)), it is indicative of correlations coming from quenching the
system into a different phase. It remains to explain the mechanism by which this happens. In this context, note that the change of ground-state energy $\Delta F$ remains extensive
also for quenches across the quantum critical point $\theta_c=0$, where $\Delta F/N$ also takes on its minimum (Fig. \ref{fig8}(b)).

Summarizing this section, we have shown that the average work and irreversible work
associated with a sudden quench of the magnetic flux across the quantum critical point $\theta_c=0$ faithfully signals the QPT,
with both quantities displaying a jump at $\theta_c=0$.
It is interesting to compare this finding to that in Ref. \cite{Campbell} where work statistics after a quantum quench was also employed to probe
equilibrium criticality, but in the Lipkin-Meshkov-Glick model. Whereas the irreversible work also there signalled a QPT when quenching across the critical point,
different from our result the average work showed no sensitivity to criticality.  The reason for this difference remains to be understood. The fact that the QPT in
the Creutz model is topological while that in the Lipkin-Meshkov-Glick model is not, is not likely to explain this intriguing dissimilarity.
%
%
\\
\section{Summary \label{Conclusions}}

In this article we have studied the quench dynamics of the Creutz model \cite{Creutz} $-$ describing spinless fermions hopping on a two-leg ladder pierced by a magnetic field.
To highlight the important role of the gap-closing zero-energy modes appearing at the quantum phase transitions between the two topologically nontrivial phases of the model,
we have taken advantage of the property that the location of these modes in the Brillouin zone can be moved by tuning the hopping amplitudes. When quenching the magnetic field
(or, equivalently, the magnetic flux through a plaquette of the ladder) to one of the quantum critical points, the revival period of the Loschmidt echo in a finite-size ladder, which does not contain
the zero-energy modes is found to be multiple of that of a commensurate finite-size ladder, which does contain these modes.
As transpires from our analysis, since information propagates through the system via the wave packets of quasiparticles,
the revival times can be identified as the time instances at which quasiparticles associated with large oscillation amplitude in the mode decomposition of the Loschmidt echo are
synchronized with the zero-energy modes.

In addition, our analysis shows that
for a quench to one of the quantum critical points, a dynamical quantum phase transition of a finite-size ladder can occur only when
the system size allows for the presence of zero-energy modes, i.e. when the gap closes completely at a wave number allowed by the finite-size quantization condition.
Again, this dramatically points to the crucial role of the zero-energy modes in the quench dynamics. Whereas the
most pronounced revivals in the Loschmidt echo happen when two conditions are satisfied $-$
large oscillation amplitudes in the mode decomposition of the Loschmidt echo {\em and} the presence of zero-energy modes synchronized
with the other modes with non-negligible oscillation amplitude (provided by quenching the system exactly
to the quantum critical point where quasiparticles are massless \cite{RJHJ2017a}) $-$ the occurrence of a dynamical quantum phase transition for a quench crossing the critical point)
only needs large oscillation amplitudes with maximum possible value.
The occurrence of a dynamical quantum phase transition for a quench {\em to} one of the critical points needs both zero-energy mode {\em and} oscillation amplitudes with maximum possible value. We notice in passing that our analysis of the role of the LE in dynamical quantum phase transitions in the Creutz ladder may be extended to topological superconductors like the Kitaev chain \cite{Bermudez2010}.

We have also investigated the quench dynamics of the Creutz model by employing tools from quantum thermodynamics.
We find that different dynamics emerge when the quench is performed {\em across} a critical point as compared to a quench {\em to} a critical point
restricted to the same phase as the initial state. As expected, when quenching across a critical point, the irreversibility of the dynamics (as measured
by the irreversible work) increases significantly. This is reflected in the different scaling with particle number of the average work and irreversible work associated with a quench across a quantum critical point as compared to a quench within the same phase $-$ a relevant piece of information when using work
statistics as a diagnostic tool for pinpointing equilibrium quantum critical points.

The results obtained add substantially to the picture how equilibrium quantum phase transitions influence nonequilibrium dynamics in a quantum many-body system,
in particular how the quantum critical zero-energy modes govern Loschmidt echo revivals and the appearance of dynamical
quantum phase transitions. More results $-$ on the quench dynamics of the Creutz model as well as on other models $-$ are expected to further advance our understanding
of the intriguing connections between equilibrium and nonequilibrium many-body physics.

\section{Acknowledgments \label{Acknowledgments}}
H. J. acknowledges support from the Swedish Research Council through Grant No. 621-2014-5972. A. L. would like to thank Sharif University of Technology for financial support under grant No. G960208. M. A. M. D acknowledges financial support from the Spanish MINECO grants FIS2012- 33152, FIS2015-67411, and the CAM research consortium QUITEMAD+, Grant No. S2013/ICE-2801. The research of M. A. M. D. has been supported in part by the U.S. Army Research Office through Grant No. W911N F-14-1-0103.



%


\end{document}